\documentclass{easychair}

\usepackage[utf8]{inputenc}
\usepackage[T1]{fontenc}
\usepackage{xspace}
\usepackage{stmaryrd}
\usepackage{./why3lang}
\usepackage{./gospel}
\usepackage{ifthen}
\usepackage[nomath,not1,notext]{stix}

\usepackage{hyperref}

\newcommand{\whyml}{\textsf{WhyML}\xspace}
\newcommand{\kidml}{\textsf{KidML}\xspace}
\newcommand{\ocaml}{\textsf{OCaml}\xspace}
\newcommand{\dafny}{\textsf{Dafny}\xspace}

\newcommand{\coq}{\textsf{Coq}\xspace}

\newcommand{\cfml}{\textsf{CFML}\xspace}
\newcommand{\vocal}{\textsf{VOCaL}\xspace}

\newcommand{\why}{\textsf{Why3}\xspace}

\newcommand{\viper}{\textsf{Viper}\xspace}

\definecolor{thegray}{rgb}{0.9,0.9,0.9}
\definecolor{colorspec}{rgb}{0,0,0.797}
\definecolor{thered}{rgb}{0.797,0,0}
\definecolor{darkgreen}{rgb}{0.797,0,0}
\definecolor{theblue}{rgb}{0,0,0.797}
\definecolor{darkgray}{rgb}{0.8477,0.8477,0.8477}
\definecolor{ocaml-bg}{rgb}{0.9,0.9,0.9}
\definecolor{thegraygray}{rgb}{0.5,0.5,0.5}

\newenvironment{judge}[2][c]
{\begin{minipage}[#1]{#2}
\hrule height2pt
\centerline{\vrule width2pt height 5pt\hfill \vrule width2pt height 5pt}
\begin{minipage}{\dimexpr\textwidth-4pt-1em}}
{\end{minipage}
\centerline{\vrule width2pt height 5pt\hfill \vrule width2pt height 5pt}
\hrule height2pt
\end{minipage}}

\newcommand{\betaxbar}{\overline{\beta x}}

\newcommand{\pibar}{\overline{\pi}}
\newcommand{\xbar}{\bar{x}}

\newcommand{\exbar}{\overline{E\xbar\Rightarrow e}}

\newcommand{\expbar}{\bar{e}}

\newcommand{\alabel}[1]{\quad\text{#1}}
\newcommand{\sep}{\;\mid\;}

\newcommand{\alphabar}{\overline{\alpha}}

\newcommand{\ghost}{\texttt{ghost}}

\theoremstyle{definition}

\usepackage{./mathpartir}

\newboolean{longversion}
\setboolean{longversion}{false} 

\usepackage{tikz}
\usetikzlibrary{arrows, shapes.misc, positioning}
\tikzset{>=stealth'}

\definecolor{thegray}{rgb}{0.9,0.9,0.9}
\definecolor{colorspec}{rgb}{0,0,0.797}
\definecolor{thered}{rgb}{0.797,0,0}
\definecolor{darkgreen}{rgb}{0.797,0,0}
\definecolor{theblue}{rgb}{0,0,0.797}
\definecolor{darkgray}{rgb}{0.8477,0.8477,0.8477}
\definecolor{ocaml-bg}{rgb}{0.9,0.9,0.9}
\definecolor{thegraygray}{rgb}{0.5,0.5,0.5}

\newcommand{\cameleer}{\textsf{Cameleer}\xspace}
\newcommand{\GOSPEL}{{\textsf{GOSPEL}}\xspace}
\newcommand{\pre}{precondition\xspace}
\newcommand{\post}{postcondition\xspace}
\newcommand{\spec}{\mathcal{S}}
\newcommand{\loops}{\mathcal{I}}
\newcommand{\atts}{\mathcal{A}}
\newcommand{\kind}{\mathcal{K}}

\usepackage{mathtools}
\usepackage{tikz-cd}
\usetikzlibrary{decorations.pathmorphing}

\newcommand\xrsquigarrow[1]{%
  \mathrel{%
    \begin{tikzpicture}[%
      baseline={(current bounding box.south)}
      ]
      \node[%
      ,inner sep=.44ex
      ,align=center
      ] (tmp) {$\scriptstyle #1$};
      \path[%
      ,draw,<-
      ,decorate,decoration={%
        ,zigzag
        ,amplitude=0.7pt
        ,segment length=1.2mm,pre length=3.5pt
      }
      ]
      (tmp.south east) -- (tmp.south west);
    \end{tikzpicture}
  }
}

\newcommand{\translate}[3]
{#1\xrsquigarrow{#2}#3}

\newcommand{\myparagraph}[1]{\vspace{-12.5pt}\paragraph{#1}}
\newcommand{\mysection}[1]{\vspace{-2pt}\section{#1}}

\title{\cameleer: a Deductive Verification Tool for OCaml\\(extended version)%
  \thanks{This work is partly supported by the HORIZON 2020 Cameleer project
    (Marie Skłodowska-Curie grant agreement ID:897873) and NOVA LINCS
    (Ref. UIDB/04516/2020)}}

\author{Mário Pereira\and António Ravara}
\authorrunning{Pereira and Ravara}
\titlerunning{\cameleer: a Deductive Verification Tool for \ocaml}

\institute{
  NOVA LINCS \& DI, Nova School of Science and Technology, Portugal}

\begin{document}

\maketitle

\begin{abstract}
  \ocaml is particularly well-fitted for formal verification. On one hand, it is
  a multi-paradigm language with a well-defined semantics, allowing one to write
  clean, concise, type-safe, and efficient code. On the other hand, it is a
  language of choice for the implementation of sensible software, \emph{e.g.},
  industrial compilers, proof assistants, and automated solvers. Yet, with the
  notable exception of some interactive tools, formal verification has been
  seldom applied to \ocaml-written programs.
  In this paper, we present the
  ongoing project \cameleer, aiming for the development of a 
  deductive verification tool for \ocaml, with a clear focus on proof
  automation. We leverage on the recently proposed \GOSPEL, \emph{G}eneric
  \emph{O}Caml \emph{SPE}cification \emph{L}anguage, to attach rigorous, yet
  readable, behavioral specification to \ocaml code. The formally-specified
  program is fed to our toolchain, which translates it into an equivalent
  program in \whyml, the programming and specification language of the \why
  verification framework. Finally, \why is used to compute verification
  conditions for the generated program, which can be discharged by off-the-shelf
  SMT solvers.
  We present successful applications of the Cameleer tool to prove functional
  correctness of several significant case studies, like FIFO queues (ephemeral
  and applicative implementations) and leftist heaps, issued from existing
  \ocaml libraries.
\end{abstract}


\mysection{Introduction}
\label{sec:introduction}

Over the past decades, we have witnessed a tremendous development in the field
of deductive software verification~\cite{filliatre11sttt}. Interactive proof
assistants have evolved from obscure and mysterious tools into \textit{de facto}
standards for proving industrial-size software projects. Notable examples
include the Sel4 verified operating system kernel~\cite{sel4}, and the verified
compilers CompCert~\cite{DBLP:conf/popl/JourdanLBLP15} and
CakeML~\cite{DBLP:journals/jfp/TanMKFON19}. On the other end of the spectrum,
the so-called \emph{SMT revolution} and the development of reusable intermediate
verification infrastructures contributed decisively to the development of
practical automated deductive verifiers. Remarkable applications of automated
verification tools include the verified version of the Microsoft's
Hypervisor~\cite{BeckertMoskal2010} and, more recently, the use of ghost
monitors~\cite{10.1145/3371070} to analyze installation scenarios of the Debian
distribution~\cite{DBLP:conf/tacas/BeckerJMRST20}.

\ifthenelse{\boolean{longversion}}
{
Interestingly, much effort has been put on the development of verification
frameworks for programs written in languages which are, by construction, hard to
get right. Languages featuring complex memory manipulation or lacking the static
guarantees of a type system, are not what one would have in mind when first
approaching the difficult task of building error-free software. On the other
hand, one can chose to conduct software verification in proof-aware languages.
These feature, by definition, the proper environment for the construction of
reliable programs, for instance, by employing expressive type systems and rich
reasoning program logics. However, such an approach suffers from the shortcoming
that it demands programmers to gain expertise in theorem proving and
verification languages. This is a natural obstacle to a wider adoption of formal
verification, since it drags the lambda-programmer away from her daily
programming tools of choice. Taking the example of the \ocaml language, despite
being a language well-fitted for verification, despite all the advances in
deductive verification and proof automation, the community still misses an easy
to use framework for the specification and verification of \ocaml code.
}

Despite all the advances in deductive verification and proof automation, little
attention has been given to the family of \emph{functional
  languages}~\cite{DBLP:conf/mpc/Regis-GianasP08}. Taking the example of the
\ocaml language, if it is a language well-fitted for verification (given its
well-defined semantics, clear syntax, and state-of-the-art type system), the
community still misses an easy to use framework for the specification and
verification of \ocaml code.

In this paper, we present \cameleer, a tool for the deductive verification of
programs directly written in \ocaml, with a clear focus on proof
automation. \cameleer uses the recently proposed
\GOSPEL~\cite{DBLP:conf/fm/ChargueraudFLP19}, a specification language for the
\ocaml language. We believe this is one of the strengths of our approach:
firstly, \GOSPEL makes a certain number of design choices, that turn it into a
clean and digestible specification language; secondly, \GOSPEL terms are written
in a subset of the \ocaml language. In the scope of this work, we have also
extended \GOSPEL to include implementation primitives, such as loop invariants
and ghost code, evolving the language from an interface specification language
into a more mature proof tool.

\cameleer takes as input an \ocaml program annotated with \GOSPEL specification
and translates it into an equivalent counterpart in \whyml, the programming and
specification language of the \why
framework~\cite{DBLP:conf/esop/FilliatreP13}. \why is a toolset for the
deductive verification of software, clearly oriented towards automated proof. A
distinctive feature of \why is that it can interface with several different
off-the-shelf theorem provers, namely SMT solvers, which greatly increases proof
automation. We believe that proof automation is another strong point of
\cameleer, on that can ease its adoption by regular \ocaml programmers. In this
paper, we present some case studies of automatically verified \ocaml modules
with our tool.

\myparagraph{Contributions.} To the best of our knowledge, \cameleer is the
first deductive verification tool for annotated \ocaml programs. It handles a
realistic subset of the language, as demonstrated by a comprehensive set of case
studies. Another contribution of this paper is our translation of the \ocaml
module language into \whyml. While sharing many common syntactic constructions,
\ocaml and \whyml differ more significantly when it comes to their module
systems. This poses interesting challenges to our translation scheme.

\myparagraph{Paper structure.}
This paper is organized as follows. Sec.~\ref{sec:warmup-example} presents a
simple example of a verified \ocaml program, intended as a smooth introduction
to the \cameleer tool and \GOSPEL language. Sec.~\ref{sec:methodology} describes
the \ocaml to \whyml translation mechanism implemented in the core of
\cameleer. Sec.~\ref{sec:case-studies} reports on relevant case studies verified
with \cameleer, including ephemeral and functorial data structures. Finally, we
present some related work in Sec.~\ref{sec:related-work} and conclude with
future work in Sec.~\ref{sec:conclusion}. The source code of \cameleer and
verified case studies are publicly available online on the \textsf{GitHub}
repository of the project\footnote{https://github.com/mariojppereira/cameleer}.

\mysection{Warmup Example}
\label{sec:warmup-example}

\ifthenelse{\boolean{longversion}}
{In this section, we present the formally verified development, in \cameleer, of
two \ocaml programs. The first one is an implementation of the Fibonacci
function; the second one is the classic binary search algorithm on an array of
integer values. The former features a purely applicative implementation, while
the latter is an example of verified code producing side-effects. This section
is intended as a gentle introduction to the \cameleer framework, the \GOSPEL
specification language, and certain important verification concepts such as
ghost code, loop invariants, and function contracts. It is intentionally written
in a rather slow pass, to provide the \ocaml programmer an enjoyable first
contact with deductive software verification.
}

In this section, we present the verified implementation of the Fibonacci
function. We first present a purely applicative version of the mathematical
definition, and then an efficient implementation using side-effects. This
section is intended as a gentle introduction to the \cameleer framework, the
\GOSPEL specification language, and certain important verification concepts such
as ghost code, loop invariants, and function contracts.
\ifthenelse{\boolean{longversion}}{It is intentionally written in a rather slow
  pass, to provide the \ocaml programmer an enjoyable first contact with
  deductive software verification.}


\myparagraph{A logical definition.} The classical mathematical definition can be
readily translated, in \ocaml, into the following recursive function:
\begin{gospel}
  let rec fib n =
    if n <= 1 then n else fib (n-1) + fib (n-2)
\end{gospel}
Though very elegant and concise, the above definition presents some pitfalls and
raises a number of questions: is this a \emph{total} function (\emph{i.e.}, it
terminates and is defined for every integer argument) and could we avoid
repeated computations in recursive calls? We focus now on the former and shall
address the latter in a moment.

The Fibonacci function is, traditionally, only defined for a \emph{non-negative}
argument and this is exactly the definition we follow here. Such a constraint on
a function arguments is a \emph{\pre}, \emph{i.e.}, it limits the range of
values the function arguments can take. We shall take this opportunity to show a
first piece of \ocaml code annotated with \GOSPEL specification:
\begin{gospel}
  let rec fib n =
    if n <= 1 then n else fib (n-1) + fib (n-2)
  (*@ requires n >= 0 *)
\end{gospel}
A \GOSPEL specification is attached to the end of a function definition and is
given within special comments of the form \of{(*@ ... *)}. The \of{requires}
clause is used to state the \pre of a function. In order to prove that every
call to \of{fib} halts, we must provide a \emph{variant} that strictly decreases
at each recursive call and has a lower-bound. Here, the value of~\of{n}
decreases at each recursive call and is bounded from below thanks to the
precondition. In \GOSPEL, we express the variant
of a function as follows:
\begin{gospel}
  let rec fib n = ...
  (*@ requires n >= 0
      variant  n *)
\end{gospel}
The given precondition and the variant form what we call the \emph{function's
  contract}. Other than serving as rigorous documentation of the function's
behavior, the interest of providing a contract is to be able to \emph{formally}
prove that the code respects the given specification. This is where the
\cameleer toolchain enters the scene.

Assuming function~\of{fib} is contained in the \ocaml file
\texttt{fibonacci.ml}, starting a proof is as easy as typing
\verb+cameleer fibonacci.ml+ in a terminal. \cameleer translates the input
program into an equivalent \whyml counterpart and launches the \why graphical
integrated development environment. This allows us to visually inspect the
verification conditions (VCs) generated by \why for the \of{fib} function: two
of them state the variant decreases and is not a negative value, at each
recursive call; the other two state the \pre holds for each recursive call. All
of these are easily discharged using SMT solvers, \emph{e.g.},
Alt-Ergo~\cite{conchon:hal-01960203}, CVC4~\cite{BCD+11}, or Z3~\cite{z3}.

\ifthenelse{\boolean{longversion}}{
It calls the \texttt{ide} sub-command of the \why tool-set, meaning we will be
conducting the proof using \why's graphical integrated development
environment. The \why framework recognizes that this is an \ocaml implementation
file (given the~\texttt{.ml} extension), hence it calls the \cameleer tool to
translate the input program into an equivalent \whyml one.



\why generates four Verification Conditions (VCs) that express the~\of{fib}
function respects the given specification: two of them state the variant
decreases and is not a negative value, at each recursive call; the other two
state the validity of the precondition, one for each recursive call. Since we
are using \why to conduct proofs, we can use several SMT solvers to try to prove
the generated verification conditions. For the above example, the
Alt-Ergo~\cite{conchon:hal-01960203}, CVC4~\cite{BCD+11}, and Z3~\cite{z3}
solvers are all able to discharge the four conditions in less than a second.}

The provided ~\of{fib} function is a naive implementation, as it
unnecessarily repeats most intermediate computations. Actually, this is not
meant to be used as an \emph{executable} implementation, but rather as a
\emph{logical} description of the Fibonacci definition. This means~\of{fib}
works more as mathematical function than a programming
one. 
In order to instruct \cameleer to consider~\of{fib} a logical
function, we decorate it with the \ocaml attribute \of{[@logic]}, as follows:
\begin{gospel}
  let [@logic] rec fib n = ...
\end{gospel}
\ifthenelse{\boolean{longversion}}{
The translation into \whyml produces the following axiom:
\begin{gospel}
  function fib int : int

  fib'def :
    forall n:int.
     n >= 0 ->
     (if n = 0 then fib n = 0
      else if n = 1 then fib n = 1 else fib n = (fib (n - 1) + fib (n - 2)))
\end{gospel}}
Under this setting, the~\of{fib} function can now be used both inside
specification clauses, as well as in regular \ocaml code.

\myparagraph{A verified efficient implementation.} Moving on to an efficient
Fibonacci function, a classical approach is to implement it as a loop that goes
from~0 to~\of{n-1}, storing in two auxiliary variables~\of{x} and~\of{y} the
values for \of{fib n} to \of{fib (n + 1)} to compute \of{fib (n + 2)}. The
\ocaml code and \GOSPEL specification are as follows:
\begin{gospel}
  let fib_imp n =
    let y = ref 0 in
    let x = ref 1 in
    for i = 0 to n - 1 do
      let aux = !y in
      y := !x; x := !x + aux
    done;
    !y
  (*@ r = fib_imp n
        requires n >= 0
        ensures  r = fib n *)
\end{gospel}
As expected, \of{fib_imp} has the same \pre as~\of{fib}. Here, we name the
result of the \of{fibonacci} function to mention it in the \emph{\post},
introduced via the \of{ensures} clause. This is exactly where we take advantage
of the fact that \of{fib} can be used as a logical function, since we state the
returned value of~\of{fib_imp n} is equal to \of{fib n}.

\ifthenelse{\boolean{longversion}}{
The presence of a \of{for} loop in the definition of \of{fib_imp} poses a
classical challenge for the deductive approach to software verification: to
supply a \emph{loop invariant}. This represents a condition that must be
satisfied at beginning and at the end of each iteration, including the first and
last iterations. One might think of a loop as a black-box, where the code that
follows will only be aware of what is computed in the loop via the
invariant. Also, one cannot expect to prove the \post of \of{fib_imp} without a
loop invariant. Coming up with a suitable loop invariant is arguably the most
difficult task in deductive verification, so let us proceed step-by-step. Since
reference~\of{y} is supposed to hold the value of~\of{fib n} at the end of
execution, we can deduce this is only true if at each iteration the value
of~\of{!y} is equal to~\of{fib i}. We write the following \GOSPEL annotation:
\begin{gospel}
  for i = 0 to n - 1 do (*@ invariant !y = fib i *) ... end
\end{gospel}
Since after the loop execution we have~\of{i = n}, indeed the supplied invariant
implies the \post. If we feed \cameleer with this program and its specification,
four verification conditions are generated: a loop invariant initialization,
which states the invariant holds before the first iteration; a loop invariant
preservation, which states the invariant holds after each iteration; two
\post{s}, one accounting for the case when the loop does not even execute
(\of{n} = 0), the other when the loop performs at least one iteration.

All of the four conditions are automatically discharged almost immediately,
except for the invariant preservation one. In fact, at each iteration,
reference~\of{y} receives the previously stored value in~\of{x} and the
invariant says nothing about the value stored in the latter reference. In order
to be able to prove that, indeed,~\of{y} always stores a value equal to
\of{fib i} we must strengthen the existing invariant as follows:
\of{ (*@ invariant ... && !x = fib (i + 1) *)}.
Now, with such a specification, all the four generated verification conditions
are automatically proved. This includes initialization and preservation for the
updated invariant.
}

As usual in deductive verification, the presence of the \of{for} loop requires
us to supply a loop invariant. Here, it boils down to
\begin{gospel}
    for i = 0 to n - 1 do
      (*@ invariant !y = fib i && !x = fib (i + 1) *)
\end{gospel}
When fed to \cameleer, four verification conditions are generated for the given
\of{fib_impl} implementation: a loop invariant initialization, which states the
invariant holds before the first iteration; a loop invariant preservation, which
states the invariant holds after each iteration; two \post, one accounting for
the case when the loop does not even execute (\of{n} = 0), the other when the
loop performs at least one iteration. All of them are automatically
discharged. The complete \ocaml implementation of the Fibonacci implementation
can be found online, at the project's GitHub
repository\footnote{\url{https://github.com/mariojppereira/cameleer/tree/master/examples/fibonacci.ml}}.

\myparagraph{The tale of the ghost code.} The \of{fib_imp} function is a
provably correct implementation of the Fibonacci function. After finishing the
proof, the specification has no computational interest and should no be part of
compiled code. This includes the \of{fib} function which is only useful as a
logical definition. This duality between parts of the code that are compiled and
other that have only proof interest is a common trait of deductive verification,
commonly known as \emph{ghost code}~\cite{gondelman16fmsd}. Some parts of a
program are marked with a special \emph{ghost} status and should be erased from
the \emph{regular} code after completing the proof effort. In \cameleer, we use
the \of{[@ghost]} attribute in order to change the status of some functions. For
the example of the \of{fib} definition, this is as simple as
\of{let [@logic] [@ghost] rec fib n}.
Building a sound mechanism for ghost code erasure is far from a trivial task,
especially in the presence of effectful
computations~\cite{parreirapereira:tel-01980343}. In
Sec.~\ref{sec:interaction-with-why}, we discuss several solutions to deal with
ghost code that we could put into practice in the scope of the \cameleer
project.

\ifthenelse{\boolean{longversion}}{
In order to obtain the \ocaml program that does not contain the implementation
of \of{fib}, we run \cameleer with the \texttt{extract} option, as follows: %
\verb+cameleer --extract fibonacci.ml+.
This instructs \cameleer to translate the \texttt{fibonacci} program into \whyml
but, instead of proceeding to the proof session, it erases any trace of ghost
code from the code and translates it back into \ocaml. To this end, we use the
\why extraction mechanism, which we developed in previous
work~\cite{parreirapereira:tel-01980343}.}

\ifthenelse{\boolean{longversion}}{
\subsection{Binary Search}
\label{sec:binary-search}

Our second example is the formal verification of the classic %
\emph{binary search} algorithm. The implementation is as follows:
\begin{gospel}
  let binary_search a v =
    let l = ref 0 in
    let u = ref (Array.length a - 1) in
    let exception Found of int in
    try while !l <= !u do
        let m = !l + (!u - !l) / 2 in
        if a.(m) < v      then l := m + 1
        else if a.(m) > v then u := m - 1
        else              raise (Found m)
      done;
      raise Not_found
    with Found i -> i
\end{gospel}
We begin our search with the upper (reference~\of{u}) and lower
(reference~\of{l}) limits of the search space set to the bounds of the whole
array. At each iteration, reference~\of{m} stores the midpoint of our
search. The subsequent case analysis is rather straightforward: (i) if the
\texttt{m}-th cell of~\of{a} contains a value less than~\of{v} (the element we
are looking for), then we update the lower limit of our search to~\of{m + 1};
(ii) if~\of{a.(m)} is greater than~\of{v}, then it is the upper that we need to
update; (iii) otherwise, we raise the locally-defined exception \texttt{Found}
with argument~\of{m} to signal we have found~\of{v} at such position. If we get
to a point after the \texttt{while} loop, it means~\of{v} cannot be found
in~\of{a}, hence we raise the \texttt{Not\_found} exception from the \ocaml
standard library.

A binary search routine only works correctly when applied to a sorted array. We
provide such an information as a \pre of \texttt{binary\_search}, using the
following auxiliary predicate:
\begin{gospel}
  (*@ predicate is_sorted (a: int array) =
        forall i1 i2. 0 <= i1 <= i2 < Array.length a -> a.(i1) <= a.(i2) *)

  let binary_search a v = ...
  (*@ i = binary_search a v
        requires is_sorted a *)
\end{gospel}
This predicate ensures the argument is a sorted array of integers if for every
two indices~\of{i1} and~\of{i2}, such that~\of{i1} is not greater than~\of{i2},
the value stored at the~\of{i1}-th cell is no greater than the value stored at
the~\of{i2}-th cell. Other than predicates, one can use \GOSPEL annotations to
introduce purely logical functions, axioms, and lemmas.

In order to prove correct our binary search implementation, we must specify the
behavior of the function on both cases. When it raises \texttt{Not\_found}
(which case we refer to as an \emph{exceptional \post}, value~\of{v} cannot be
found in any cell within the bounds of~\of{a}; if \texttt{binary\_search} does
not raise any exception to its outer scope (note that \texttt{Found} is cached
inside the body of \texttt{binary\_search}), then~\of{v} is found at the index
returned by the function, which we named~\of{i} in our specification. We update
our \GOSPEL specification as follows:
\begin{gospel}
  (*@ i = binary_search a v
      ...
      raises   Not_found -> forall i. 0 <= i < Array.length a -> a.(i) <> v
      ensures  0 <= i < Array.length a && a.(i) = v *)
\end{gospel}
If we feed this program and specification to \cameleer, 6 VCs are generated: one
for the termination of the loop; one to check that we do not divide by zero; two
to prove that we only access the array in bounds; and finally one VC for each
\post. At this point, we are only able to prove one of them, namely the absence
of division by zero. This comes at no surprise: the presence of a \texttt{while}
loop in the definition of \of{binary_search} poses a classical challenge for the
deductive verification approach, \emph{i.e.}, to provide a loop specification in
the form of a \emph{loop variant} and a \emph{loop invariant}. The former, as
for recursive definitions, is used to specify termination of a computation. In
the case of \of{binary_search}, this is as simple as
\begin{gospel}
        while !l <= !u do
        (*@ variant !u - !l *)
\end{gospel}
At each iteration, either the value of~\of{u} decreases or the value of~\of{l}
increases, hence \texttt{!u - !l} is a valid termination measure. This is
sufficient to prove the loop always terminates, for any input array~\of{a}.

The loop invariant part is a bit more challenging. A loop invariant represents a
condition that must be satisfied at beginning and at the end of each
iteration. One might think of a loop as a black-box, hence the code that follows
will only be aware of what is computed in the loop via the invariant. Hence, we
cannot expect to prove the \post of \of{binary_search} without a loop invariant,
since after execution we know nothing about the argument of \texttt{Found}.
Coming up with a suitable loop invariant is arguably the most difficult task in
deductive verification, so let us proceed step-by-step. First, we must
constraint the values references~\of{u} and~\of{l} can store. In our
example,~\of{l} is always non-negative, while~\of{u} is always strictly smaller
than the length of the array. This is stated as follows:
\begin{gospel}
        while !l <= !u do
        (*@ variant   !u - !l *)
        (*@ invariant 0 <= !l && !u < Array.length a *)
\end{gospel}
With such an invariant at hand, we are already able to prove every generated VC,
except for the exceptional \post.

In order to prove the last remaining VC, we must strengthen our invariant with a
condition that states if the value~\of{v} is in~\of{a} in a certain cell~\of{i},
then it must be the case that~\of{i} is always contained within~\of{!l}
and~\of{!u}. We update our invariant as follows:
\begin{gospel}
        (*@ invariant 0 <= !l && !u < Array.length a *)
        (*@ invariant forall i. 0 <= i < Array.length a -> a.(i) = v ->
            !l <= i <= !u *)
\end{gospel}
When we raise the \of{Not_found} exception, we are after the loop execution
hence we have in our proof context the condition~\of{!l > !u}. Together with the
loop invariant, we apply the following \emph{contra-positive} reasoning: since
it is the case that \texttt{!l > !u}, then it also the case that there exists no
index~\of{i} of the array such that~\of{a.(i) = v} (according to the
invariant). This completes the proof of the exceptional \post.

With the strengthened invariant, all the VCs generated for
\texttt{binary\_search} are automatically discharged using a combination of the
Alt-Ergo and CVC4 SMT solvers. The complete \ocaml development and \GOSPEL
specification for the \texttt{binary\_search} function can be found online, at
the project's GitHub
repository\footnote{\url{https://github.com/mariojppereira/cameleer/tree/master/examples/binary_search.ml}}.


}
\mysection{Methodology}
\label{sec:methodology}

This section gives an overview on the core of the \cameleer tool, from the
\ocaml code annotated with \GOSPEL specification to the generation of an
equivalent \whyml program. In Sec.~\ref{sec:ocaml-compiler}, we describe how we
use the \GOSPEL toolchain to attach specification to certain nodes in the
\ocaml AST. In Sec.~\ref{sec:transl-into-whyml} we define our \ocaml to \whyml
translation as a set of inference rules. Finally,
Sec.~\ref{sec:interaction-with-why} explains how we are currently using \why as
an intermediate verification framework.

\subsection{Using \GOSPEL toolchain}
\label{sec:ocaml-compiler}

The \cameleer tool relies on the use of the \GOSPEL toolchain%
\footnote{\url{https://github.com/vocal-project/vocal}} to parse and manipulate
the \ocaml abstract syntax tree. It provides a patched version of the \ocaml
parser that recognizes \GOSPEL special comments and converts them to regular
\ocaml attributes. For instance, the \texttt{fib\_imp} specification from
Sec.~\ref{sec:warmup-example}
\ifthenelse{\boolean{longversion}}{\begin{gospel}
  let fib_imp n = ...
  (*@ r = fib_imp n
      requires n >= 0
      ensures  r = fib n *)
\end{gospel}}
is translated into the following post-item attribute:
\begin{gospel}
  let fib_imp n = ...
  [@@gospel ``r = fib_main n requires n >= 0 ensures r = fib n'']
\end{gospel}
The payload of a \GOSPEL attribute is, hence, a string that contains the
user-supplied specification\footnote{In fact, \GOSPEL elements can be directly
  provided using attributes in the source code.}.
\ifthenelse{\boolean{longversion}}{
This is not different from how
the \ocaml parser treats documentation comments of the form \texttt{(** $\ldots$
  *)}.}
\ifthenelse{\boolean{longversion}}{
Finally, \GOSPEL elements that are not attached to any \ocaml item are
translated into \emph{floating} attributes. For instance, the predicate
\texttt{is\_sorted} from Sec.~\ref{sec:binary-search} is translated into
\texttt{[@@@gospel ``predicate is\_sorted ...'']}}
The \GOSPEL attributes are processed by a dedicated parser and
type-checker~\cite{DBLP:conf/fm/ChargueraudFLP19}, where specifications are
attached to nodes of a patched version of the \ocaml AST.
\ifthenelse{\boolean{longversion}} {For instance, the following represents the
  case of an \ocaml anonymous function with a \GOSPEL specification attached:
\begin{gospel}
  type expression_desc =
    ...
    | Sexp_fun of
        arg_label * expression option * pattern * expression * fun_spec option
\end{gospel}
} This custom AST is the entry-point for our \ocaml to \whyml translation, which
we describe next.

\subsection{Translation into \whyml}
\label{sec:transl-into-whyml}

We present our translation from \ocaml to \whyml as set of inference rules. All
the auxiliary functions and predicates are total definitions. We focus here on a
subset of the \ocaml and \whyml languages. The complete definitions are depicted
in Fig.~\ref{fig:core-ocaml} and Fig.~\ref{fig:whyml}, respectively. On the
\whyml side, we omit the definition of~$t$ which stands for the logical subset
of \whyml. For a comprehensive definition of this part of \whyml, we refer the
reader to the \why reference manual~\cite[Chap. 7]{why3manual}. We do not intend
to use this section as an heavy formalization of our translation, but rather as
a comprehensive presentation of the \ocaml subset that \cameleer can
handle. \cameleer will report a dedicated error message if a user tries to
translate an \ocaml program that syntactically falls out of the supported
fragment. It is worth noting that our translation is purely syntactic, as we
build on the \GOSPEL toolchain which is based on a PPX approach. In particular,
this means that typing the translated \ocaml program is left as a task to \why.
Making our translation type-directed, or at least type-aware, is left as future
work.

\ifthenelse{\boolean{longversion}}
{
  These definitions are very close to the ASTs issued by the \ocaml
  and \whyml parsers. The \whyml formalization is also strongly inspired by the
  \kidml language, presented in Pereira's PhD
  thesis~\cite{parreirapereira:tel-01980343}.
}

\begin{figure}
  \centering
\[
  \begin{array}{llr}
    e & ::= \; x \sep \nu
        \sep \mathtt{if}\: e\: \mathtt{then}\: e\: \mathtt{else}\: e
        \sep \mathtt{match}\; e\; \mathtt{with}\;
        \overline{\talloblong\,p\Rightarrow e}
    & \alabel{Expressions}\\
      & \sep \mathtt{let} \: \atts \; x = e \; \atts \; \mathtt{in} \; e \\
      &  \sep \mathtt{let} \: \atts \; \mathtt{rec} \: f = e \; \atts \;
        \overline{\mathtt{and}\;f = e \; \atts} \; \mathtt{in} \; e
        \sep f\,\expbar \\
      & \sep \{ \overline{f = e} \} \sep e.f \sep e.f \leftarrow e \\
      & \sep \mathtt{raise} \; E\expbar
        \sep \mathtt{try} \; e \; \mathtt{with} \; \exbar \\
      & \sep \mathtt{while} \: e \: \mathtt{do} \: \atts \: e \: \mathtt{done}
        \sep \mathtt{assert\;false} \\[.5em]

    p & ::= \; \_ \sep x \sep \overline{p} \sep C (p)
        \sep p\: \talloblong\: p \sep p\;\mathtt{as}\;x \sep p : \tau
        \sep \mathtt{exception}\;p & \alabel{Patterns}\\[.5em]

    \nu & ::= \; n \sep \mathtt{true} \sep \mathtt{false} \sep
          \mathtt{fun}\:\atts\:(x\beta) \rightarrow e & \alabel{Values}\\[.5em]

    \beta & ::= \; \texttt{reg} \sep \texttt{ghost} & \alabel{Ghost attribute}
    \\[.5em]

    \tau & ::= \; \alpha \sep \tau\rightarrow\tau \sep \overline{\tau}
           \sep \overline{\tau}\:C & \alabel{Type expression} \\[.5em]

    \pi & ::= \; \beta\tau & \hspace{-40pt}
                             \alabel{Type with ghost status}\\[.5em]

    d & ::= \; \texttt{exception}\: E : \pibar
        \sep \texttt{type} \; td  \; \overline{\texttt{and}\; td}
    & \alabel{Top-level declarations} \\
      & \sep \mathtt{let}\: \atts \; f = e \: \atts
        \sep \mathtt{let}\: \atts \; \texttt{rec}\; f = e \: \atts\;
        \overline{\mathtt{and}\;f = e \; \atts}\\
      & \sep \texttt{module}\:\mathcal{M} = m\\[.5em]

    td & ::= \; \alphabar \: T
         \sep \alphabar \: T = \tau
         \sep \alphabar \: T = \{\: \overline{f : \pi}\:\}\: \atts
         \sep \alphabar \: T =
         \overline{\talloblong\,C\,\texttt{of}\,\overline{\tau}}
    & \alabel{Type definition}\\[.5em]

    m & ::= \; \texttt{struct}~~\overline{d}~~\texttt{end}
        \sep \texttt{functor}(\mathcal{X}: mt) \rightarrow m
            & \alabel{Modules} \\[.5em]

    mt & ::= \; \texttt{sig}\;\overline{s}\;\texttt{end} & \alabel{Module
                                                           types}\\[.5em]

    s & ::= \; \texttt{val}\:\atts\;f : \pi \: \atts
        \sep \texttt{type}\: td \; \overline{\texttt{and}\; td}
    & \alabel{Signatures}\\[.5em]

    p & ::= \; \overline{d} & \alabel{Program}
  \end{array}
\]
\caption{Syntax of core \ocaml.}
\label{fig:core-ocaml}
\end{figure}

\begin{figure}[h!]
  \centering
\[
  \begin{array}{llr}
    e & ::= \; x \sep \nu
        \sep \mathtt{if}\: e\: \mathtt{then}\: e\: \mathtt{else}\: e
        \sep \mathtt{match}\; e\; \mathtt{with}\;
        \overline{\talloblong\,p\Rightarrow e} \;
        \mathtt{end}
    & \alabel{Expressions}\\
      & \sep \mathtt{let} \; \kind\; \beta x = e \; \mathtt{in} \; e \\
      & \sep \mathtt{rec}\;\kind\;f(\overline{\beta x})\;\spec = e \;
        \overline{\mathtt{with}\;\kind\;f (\overline{\beta x}) = e \; \spec} \;
        \mathtt{in} \; e
        \sep f\,\expbar \\
      & \sep \{ \overline{f = e} \} \sep e.f \sep e.f \leftarrow e \\
      & \sep \mathtt{raise} \; E\expbar
        \sep \mathtt{try} \; e \; \mathtt{with} \; \exbar \; \mathtt{end} \\
      & \sep \mathtt{while} \: e \: \mathtt{do} \: \loops \: e \: \mathtt{done}
        \sep \mathtt{absurd}
        \sep \mathtt{ghost} \; e\\[.5em]

    p & ::= \; \_ \sep x \sep \overline{p} \sep C p
        \sep p\: \talloblong\: p \sep p\;\mathtt{as}\;x \sep p : \tau
        \sep \mathtt{exception}\;p & \alabel{Patterns}\\[.5em]

    \nu & ::= \; n \sep \mathtt{true} \sep \mathtt{false}
          \sep \mathtt{fun}\;\kind\; (\overline{\beta x}) \; \spec \rightarrow
          e & \alabel{Values} \\[.5em]

    \beta & ::= \; \texttt{reg} \sep \texttt{ghost} & \alabel{Ghost status}
    \\[.5em]

    \tau & ::= \; \alpha \sep \tau\rightarrow\tau \sep \overline{\tau}
           \sep C\,\overline{\tau} & \alabel{Type expression} \\[.5em]

    \pi & ::= \;\beta\tau & \hspace{-4pt}\alabel{Type with ghost status}\\[.5em]

    \kind & ::= \;\texttt{reg} \sep \texttt{logic} & \alabel{Function
                                                          kind}\\[.5em]

    \spec & ::= \; \texttt{requires}\; \overline{t}~~
             \texttt{ensures}\; \overline{t}~~
            \texttt{variant}\; \overline{t}
    & \alabel{Function specification} \\[.5em]

    \loops & ::= \; \texttt{invariant}\; \overline{t}~~\texttt{variant}\;
             \overline{t} & \alabel{Loop specification} \\[.5em]

    d & ::= \; \texttt{exception}\: E : \pibar
        \sep \texttt{type}\: td \; \overline{\texttt{with}\:td}~~~~~
                                  & \alabel{Top-level declarations} \\
      & \sep \mathtt{let} \; \mathcal{K} \; f = e \\
      & \sep \mathtt{let} \; \texttt{rec} \; \mathcal{K} \; f(\overline{\beta x})
        \:\spec = e \;
        \overline{\mathtt{with}\;\kind\;f (\overline{\beta x})\:\spec = e}\\
      & \sep \mathtt{val}\;\mathcal{K}\;\beta f(\overline{x:\pi})\:\spec : \pi
        \sep \texttt{scope} \;\mathcal{M}\;\overline{d} \; \texttt{end} \\[.5em]

    td & ::= \; T\alphabar
         \sep T\alphabar = \tau
         \sep
         T\alphabar = \{\: \overline{f : \pi}\:\}\:\texttt{invariant}\:\bar{t}
         \sep T\alphabar = \overline{\talloblong\,C\,\overline{\tau}}
    & \alabel{Type definition}\\[.5em]

    p & ::= \; \texttt{module}\;\mathcal{M}\;\overline{d}\;\texttt{end}
    & \alabel{Program}
  \end{array}
\]
\caption{Syntax of core \whyml.}
\label{fig:whyml}
\end{figure}

\myparagraph{Expressions.} Selected \ocaml expressions include variables ($x$
ranges over program variables, while~$f$ is used for function names), the
conditional \texttt{if..then..else}, local bindings of (possibly recursive)
expressions, function application, records manipulation (for simplicity, we
assume every field to be mutable), treatment of exceptions, loop construction,
and finally the \texttt{assert false} expression. Values include numerical and
Boolean constants, as well as anonymous functions where arguments are annotated
with a ghost status. We only consider functions as valid recursive definitions
and application is limited to the application of a function name to a list of
arguments. The latter is just to ease our presentation; the former is due to
recursive definitions in \whyml being limited to functions. Finally, $\atts$
notation stands for a (possibly empty) placeholder of \ocaml attributes,
representing the original place in the expression where \GOSPEL elements are
introduced. For instance, the first $\atts$ in a \texttt{let..rec} expression
can contain the \of{[@ghost]} and \of{[@logic]} attribute, while the second one
stands for the function specification. We omit the definition of~$\tau$ and~$t$,
respectively from the \ocaml and the \whyml sides. The former stands for the
grammar of \ocaml types, while the latter is the logical subset of \whyml.


The \ocaml and \whyml languages are very similar, hence our translation of
expressions is mostly an isomorphism. We give the complete set of rules in
Appendix~\ref{sec:transl-expr} and explain here only the more subtle aspects of
this translation. Let us begin with the translation of an \ocaml expression of
the form \texttt{let..rec..in} into its \whyml counterpart. The corresponding
translation rule is the following:
\[
  \inferrule*[Left=(ERec)]
  {
    \translate{e_0}{function}{\overline{\beta x},e_0'} \\
    \translate{e_1}{expression}{e_1'} \\\\
    \neg\textit{is\_ghost}(\atts_0) \\
    \mathit{kind}(\atts_0) = \kind \\
    \translate{\atts_1}{\textit{function spec}}{\spec_1}}
  {\translate{\mathtt{let} \: \atts_0 \; \mathtt{rec} \: f_0 = e_0 \;
      \atts_1 \; \mathtt{in} \; e_1}{expression}
    {\mathtt{rec}\;\kind\;f(\overline{\beta x})\;\spec_1 = e_0' \; \mathtt{in}
      \; e_1'}}
\]
For the sack of presentation, we use here only a single recursive function and
omit any mutually recursive definition. In fact, translating a set of mutual
definitions simply amounts to a recursive call to our expressions translation
procedure, as depicted in Appendix~\ref{sec:transl-expr}. Let us consider the
following generic expression as a running example to explain the \textsc{(ERec)}
rule:
\begin{gospel}
  let rec foo (x [@ghost]) y = e0
  (*@ r = foo x y
        requires ... variant  ... ensures  ... *)
  in e1
\end{gospel}
The second premise of the rule translates expression~\of{e1} and it simply
amounts to a recursive call to the translation scheme. The first premise is a
bit more evolving, hence we explain it in more detail.
The definition of \of{foo} is de-sugared by the \ocaml parser into the following
Curried expression:
\begin{gospel}
  let rec foo = fun (x [@ghost]) -> fun y -> e
\end{gospel}
When translating into \whyml, we revert such an operation: we traverse the body
of \of{foo}, building a list of ghost-annotated arguments from the argument of
each~\texttt{fun} construction. The body of the translated function is the body
of the last~\texttt{fun}. This is done in the first premise of the rule, using
the $\translate{}{function}{}$ operation. This conversion into multi-argument
functions is justified by the limits of \why when it comes to higher-order and
anonymous functions. In \whyml, one can only define \emph{pure} anonymous
functions, \emph{i.e.}, free of any side effects. Hence, directly translating
the Curried definition of \of{foo} would yield a syntactically correct \whyml
expression, however this would be rejected by the language type-and-effect
system.

The other three premises deal with specification elements. The first uses the
\textit{is\_ghost} operation to test whether the \of{[@ghost]} attribute is
provided in~$\mathcal{A}_0$. If that is the case (rule \textsc{(ERecGhost)} in
Appendix~\ref{sec:transl-expr}), the function body is translated into %
\of{ghost e}. The kind of a function is either \of{reg} (only usable as a
program function) or \of{logic} (also usable inside specification). We introduce
the $\mathit{kind}(\cdot)$ operation, which also retrieves the function kind
from~$\mathcal{A}_0$ (in case of a regular function, the attribute can be
omitted). The last premise translates the supplied \GOSPEL \pre{s}, variants,
and \post{s} into the \whyml specification language. We omit the definition of
$\translate{}{\textit{function spec}}{}$ since this is a trivial (syntactic)
transformation.

We highlight two more interesting cases of expressions translation: the
\texttt{assert false} construction and local non-recursive bindings. Contrarily
to any other \texttt{assert} expression, \texttt{assert false} is used in \ocaml
to indicate unreachable points in the code and ``is treated in a special way by
the \ocaml type-checker''\footnote{To quote a comment from the \ocaml compiler
  source code itself.}. \whyml features the \texttt{absurd} construction which
has the exact same semantics, which greatly simplifies our translation effort
(rule \textsc{(EAbsurd)} in Appendix~\ref{sec:transl-expr}). Finally, when
translating a \texttt{let..in} expression, we need to account for both the
introduction of a local function, as well as the binding of a non-functional
value. The translation rule for the latter is as follows:
\[
  \inferrule*[Left=(ELet)]
  {\neg\textit{is\_ghost}(\atts) \\ \neg\textit{is\_functional}(e_0) \\
    \translate{e_0}{expression}{e_0'} \\ \translate{e_1}{expression}{e_1'}}
  {\translate{\mathtt{let}\:\atts\:x = e_0\;\atts'\;\mathtt{in}\;e_1}
    {expression}{\mathtt{let}\: \mathtt{reg} \: x = e_0'\;\mathtt{in} \;e_1'}}
\]
This rule stands for the sub-case where the bound variable is regular, hence the
use of the \texttt{reg} kind. Any \GOSPEL specification possibly contained in
$\atts'$ is ignored. Finally, the use of the auxiliary predicate
$\textit{is\_functional}(\cdot)$ is what allows us to distinguish between
locally-bound variables and functions. This predicate decides whether~$e_0$ is a
\of{fun x -> ...} expression, in which case we introduce a \whyml local function
(rule \textsc{(ELetFun)} in Appendix~\ref{sec:transl-expr}). This approach does
not take partial applications into account, which are directly translated into
its \whyml counterpart. If a partial application introduces an effectful
computation, this will be rejected by the \why type system.

\paragraph{Top-level declarations.} Selected top-level declarations include
exceptions and type declaration, (mutually-recursive) function definition, and
introduction of sub-modules. An exception takes a list of~$\pi$ values, types
annotated with a ghost status, to account for the possibility of ghost
arguments. In \why vocabulary, this is the \emph{mask} of an
exception~\cite[Chap. 3.1]{parreirapereira:tel-01980343}. The complete set of
translation rules for top-level declarations and type definitions is given in
Appendix~\ref{sec:transl-top-level}. We highlight here the cases of record type
definition and sub-modules.

The attribute~$\atts$ after a record type definition is used to express in
\GOSPEL a \emph{type invariant}, \emph{i.e.}, a predicate that every inhabitant
of such type must satisfy. Type invariants are readily supported by \why, as
depicted in rule \textsc{(TDRecord)} in Fig.~\ref{fig:type-def},
Appendix~\ref{sec:transl-top-level}.
%
%
Each field of the record type is also annotated with a ghost status. This is a
common practice in deductive verification: some fields act as \emph{logical
  models} of the record value; these can be explored within the proof to reason
about the represented data structure. It is worth noting that in \whyml,
contrarily to \ocaml, type arguments are introduced on the right-hand side of
the type name.

Finally, translation of a sub-module definition is guided by the following rule:
\[
  \inferrule*[Left=(DModule)]
  {\translate{m}{module}{\overline{d}}}
  {\translate{\mathtt{module}\:\mathcal{M} = m}{declaration}
    {\mathtt{scope}\:\mathcal{M}\;\overline{d}\;\mathtt{end}}}
\]
We translate the module expression~$m$ into a list of \whyml
declarations. \whyml does not feature the notion of sub-module, hence we
encapsulate the translated declarations into a \emph{scope}, the \whyml unit for
namespaces management. In what follows, we provide a
more detailed account of the \whyml module system and its differences with
respect to that of \ocaml.

\myparagraph{Modules.} The most interesting cases in our translation is how we
deal with the modules language from the \ocaml side. A \whyml program is a list
of modules, a module is a list of top-level declarations, and declarations can
be organized within scopes. 

The first module expression we take into account is the \texttt{struct..end}
construction. This is translated into a \whyml declarations, as depicted in rule
\textsc{(MStruct)} (Appendix~\ref{sec:transl-module-expr}).
%
We note this does not change the structure and code organization
of the original program, since a \texttt{struct..end} expression follows a
\texttt{module} declaration. Hence, a declaration of the form
\texttt{module\,$\mathcal{M}\;\bar{d}$\;end} is translated into
\texttt{scope\,$\mathcal{M}\;\bar{d}$\;end}.

Functors are a central notion when programming in \ocaml, so it is out of
question to develop a verification tool for \ocaml without a (at least minimal)
support for functors. \whyml does not feature a syntactic construction for
functors; instead, these are represented as modules containing only abstract
symbols~\cite{paskevich20isola}. Thus, we propose the following translation
rule:
\[
  \inferrule*[Left=(MFunctor)]
  {\translate{mt}{\textit{module type}}{\overline{d}} \\
   \translate{m}{module}{\overline{d'}}}
  {\translate{\mathtt{functor}(\mathcal{X}: mt)\rightarrow m}{\mathit{module}}
    {\mathtt{scope}\:\mathcal{X}\;\overline{d}\;\mathtt{end}\;\overline{d'}}}
\]
Each \texttt{functor} expression is translated into a \whyml scope followed by a
list of declarations.
\ifthenelse{\boolean{longversion}}{
The latter is the resulting of translating the body of the functor, while the
former encapsulates the functor argument.}
For instance, the \ocaml functor
\of{module Make = functor (X: ...) -> struct ... end}, is translated into the
following \whyml excerpt: \of{scope Make scope X ... end ... end}.
The given transformation is the dual of what is actually implemented in the \why
extraction machinery: every \whyml expression of the form %
\of{scope A scope B ...} is translated into \of{module A (B: ...) ...}, \emph{as
  long as} scope~\texttt{B} features only abstract symbols.

\myparagraph{Signatures.} The argument of a functor is expressed as a
\emph{module type}, \emph{i.e.}, a \emph{signature} of the form
\texttt{sig..end}. This encapsulates a list of declarations belonging to the
\ocaml signature language, which are translated into a list of \whyml
expressions, according to rule \textsc{(\textsc{MTSig})}
(Appendix~\ref{sec:transl-module-expr}).
%
%
Contrarily to \ocaml, \whyml does not impose a separation between signature
(interface) and structure (implementation) elements. In particular, the \whyml
surface language allows one to include non-defined \texttt{val} functions and
regular \texttt{let} definitions in the same namespace. We give the following
translation rule for \texttt{val} declarations:
\[
  \inferrule*[Left=(SVal)]
  {\neg\textit{is\_ghost}(\atts) \\ \textit{kind}(\atts) = \kind \\
    \translate{\atts'}{\textit{function spec}}{\spec} \\
    \translate{\pi,\atts'}{\textit{function args}}{\overline{x:\pi'},\pi_{res}}}
  {\translate{\mathtt{val}\: \atts \: f : \pi \: \atts'}{signature}
    {\mathtt{val}\:\kind\: f (\overline{x:\pi})\:\spec : \pi_{res}}}
\]
The name of the arguments are retrieved from the function specification
(Sec.~\ref{sec:leftist-heaps} features an example of such case). Non-defined
functions can also be declared as ghost and/or logical functions. For brevity,
the case of ghost \texttt{val} is omitted. The complete set of translation rule
for signature items can be found in Appendix~\ref{sec:transl-sign}.

\myparagraph{Programs.} An \ocaml program is simply a list of top-level
declarations. These are translated into a \whyml module, as follows:
\[
  \inferrule*[Left=(Program)]
  {\translate{\overline{d}}{declaration}{\overline{d'}}}
  {\translate{\overline{d}}{program}
    {\mathtt{module}\,\mathcal{M}\;\overline{d'}\;\mathtt{end}}}
\]
The name~$\mathcal{M}$ of the generated module is issued from the \ocaml file
that contains the original program. If file \texttt{foo.ml} contains the
program~$p$, it gets translated into \texttt{module\:Foo\;$p$\;end}. In summary,
we generate a \whyml program containing a single module, which represents the
top-level module of an \ocaml file. In turn, each sub-module is translated into
a \whyml scope, with a special treatment for functorial definitions.

\subsection{Interaction with \why}
\label{sec:interaction-with-why}

\paragraph{\why front-end.} One distinguished feature of the \why architecture
is that it can be extended to accommodate new front-end
languages~\cite[Chap. 4]{why3manual}. Building on the translation scheme
presented in previous section, we use the \why API to build an in-memory
representation of the \whyml program and to register \ocaml as an admissible
input format for \why.  \ifthenelse{\boolean{longversion}} {In practice, this is
  done by calling the \texttt{register\_format} function from the \texttt{Env}
  module of the \why API, as follows:
\begin{gospel}
  register_format mlw_language "ocaml" ["ml"] read_channel ~desc:"OCaml format"
\end{gospel}
It registers a new input format named \texttt{ocaml}, which should be recognized
for files with extension \texttt{.ml}. The \texttt{read\_channel} function reads
a file name from an input channel, invokes the \GOSPEL machinery to parse the
\ocaml code and type-check \GOSPEL attributes, and finally calls our translation
function to produce a program of the \texttt{mlw\_language}. Finally, we
register the \cameleer tool as a plugin for the \why framework. This means that
}
We can use any \why tool, out of the box, to process a \texttt{.ml} file. For
instance, one could use directly the command \texttt{why3 ide bar.ml} to trigger
our \ocaml input format and to call the \why IDE on the translation of the
\texttt{bar.ml} file.
\ifthenelse{\boolean{longversion}}{Our programming effort currently amounts to
1.1K of non-blank lines of \ocaml code.}



\ifthenelse{\boolean{longversion}}{
\myparagraph{Conducting proofs in \why.}
\ifthenelse{\boolean{longversion}}
{
\begin{figure}
  \centering
  \includegraphics[scale=0.32]{img/why_ide.png}
  \caption{\why proof session for lemma \texttt{mem\_decomp}.}
  \label{fig:why3_ide}
\end{figure}
}
Invoking the \cameleer tool on an \ocaml file launches the \why Integrated
Development Environment on the generated \whyml program. On one hand one can
experiment with multiple SMT solvers to discharge different VCs. On the other
hand, one can use \emph{logical transformations} as mean to conduct lightweight
interactive proofs inside \why~\cite{dailler2018}. Consider, for instance, the
following predicate defined in \GOSPEL
\begin{gospel}
  (*@ predicate mem (x: 'a) (l: 'a list) = match l with
      | []     -> false
      | y :: r -> x = y || mem x r *)
\end{gospel}
together with a lemma about it:
\begin{gospel}
  (*@ lemma mem_decomp: forall x: 'a, l: 'a list.
        mem x l -> exists l1 l2: 'a list. l = l1 @ (x :: l2) *)
\end{gospel}
Predicate \of{mem_decomp} decides whether~\of{x} can be found in
list~\of{l}. The \of{mem_decomp} lemma states that, if indeed~\of{x} occurs
in~\of{l}, then it must be the case that~\of{l} can be decomposed in a
prefix~\of{l1} and a suffix~\of{l2} with element~\of{x} in between. When fed to
\cameleer, this \GOSPEL specification is translated into a \whyml predicate and
a \emph{goal} with the statement of \of{mem_decomp}. Such a property can be
easily proved by structural induction on the list~\of{l}, but such kind of
reasoning is out of scope for nowadays automated solvers. Nonetheless, the \why
IDE allows us to use the \texttt{induction\_arg\_ty\_lex} on~\of{l}, which
generates two new VCs: one for the base case of \of{mem_decomp}, the second one
for the inductive step, which are then easily discharged.
}

\ifthenelse{\boolean{longversion}}
{Fig.~\ref{fig:why3_ide} presents the proof of \texttt{mem\_decomp} in the \why
IDE. On the left-1hand side we can observe the application of
\texttt{induction\_arg\_ty\_lex}, as well as \texttt{introduce\_premises} (to
introduce variables and premises in the proof context) and \texttt{split\_vc}
(to split a conjunction into several clauses), together with the use of the
Alt-Ergo and CVC4 solvers. On the right-hand side, the IDE presents the current
\emph{task}, \emph{i.e.}, the list of logical declarations that are fed to
external provers. It is worth noting that these are \whyml elements, hence
syntactic difference with respect to \GOSPEL and \ocaml are expected
(\emph{e.g.}, the \texttt{match..with} expression is terminated by the
\texttt{end} keyword).}

Other than the \texttt{ide}, we could use the \texttt{extract} command to erase
any trace of ghost from the original code and print an equivalent \ocaml
implementation. This would provide us with the necessary guarantees about the
semantics and typedness of the extracted program. However, we believe a solution
that is directly integrated with the \ocaml compilation chain is more organic
and natural to the programmer. We currently working on a PPX that generates a
new \ocaml AST without ghost code, which uses the \why extraction as an internal
ingredient.


\ifthenelse{\boolean{longversion}}{
  In practice, invoking \cameleer with the \of{--extract} option on
file \texttt{foo.ml} is equivalent to directly calling
\verb+why3 extract -D cameleer.drv foo.ml+. The \texttt{cameleer.drv} file is an
extraction \emph{driver}, \emph{i.e.}, a text file that states how \whyml
symbols should be extracted back into \ocaml. For instance, the following
excerpt of the \texttt{cameleer} driver
\begin{gospel}
  module list.List
    syntax type list "
    syntax val  Nil  "[]"
    syntax val  Cons "
  end
\end{gospel}
instructs extraction to map values of type \texttt{list}, represented in \whyml
by the \texttt{Nil} and \texttt{Cons} constructors, to their counterpart in
\ocaml (the empty list \texttt{[]} and the \texttt{(::)} operator,
respectively). The \texttt{list} type is converted into \ocaml type with the
same name, however the argument is placed on the left of the type name. In an
extraction driver, \texttt{\%n} is placeholder for the \texttt{n}-th argument of
the extracted symbol.}

\myparagraph{Limitations of using \why.} \whyml and \GOSPEL are very similar
specification language. Moreover, they share some fundamental principals, namely
the arguments of functions are not-aliased by construction and each data
structure carries an implicit representation predicate. This makes the
translation from \GOSPEL to \whyml a very natural process. However, one can use
\GOSPEL to formally specify some \ocaml programs which cannot be translated into
\whyml. This is much evident when it comes to recursive ephemeral data
structures. Consider, for instance, the \texttt{cell} type definition from the
\texttt{Queue} module of the \ocaml standard
library\footnote{\url{https://caml.inria.fr/pub/docs/manual-ocaml/libref/Queue.html}}:
\begin{gospel}
  type 'a cell = Nil | Cons of { content: 'a; mutable next: 'a cell }
\end{gospel}
As we attempt to translate such data type into \whyml, we get the following
error:
\begin{verbatim}
  This field has non-pure type, it cannot be used in a recursive type definition
\end{verbatim}
Recursive mutable data types are beyond the scope of \why 's type-and-effect
discipline~\cite{gondelman16reg}. The solution would be to resort to an
axiomatic memory model of \ocaml in \why~\cite{fpds18jfla}, or to employ a
richer program logic, \emph{e.g.}, Separation Logic~\cite{10.5555/645683.664578}
or Implicit Dynamic Frames~\cite{smans2009implicit}. We leave such an extension
to the \cameleer infrastructure for future work.




\mysection{Case Studies}
\label{sec:case-studies}

\subsection{FIFO Queue}
\label{sec:queues}

\ifthenelse{\boolean{longversion}}{\myparagraph{Ephemeral implementation.}}
\ifthenelse{\boolean{longversion}}{
This is represented by the following type:
\begin{gospel}
  type 'a t = 'a list * 'a list
\end{gospel}
The first list represents the \emph{prefix} of the queue, while the second one
stands for the \emph{suffix} part. Prefix elements are stored in the correct
order and suffix elements are stored in \emph{reversed} order. Elements are
pushed into the head of the suffix list and popped off the head of prefix. This
data structure also maintains the \emph{invariant} that if the prefix is empty,
then so is the suffix. This guarantees that a queue has a constant amortized
complexity for a combination of push and pop operations.
In order to formally express the behavioral properties of the queue data
structure, we shall equip type~\of{t} with a model field and a \GOSPEL
invariant, as follows:
\begin{gospel}
  type 'a t = {
    self : 'a list * 'a list;
    view : 'a list [@ghost];
  } (*@ invariant let (prefix, xiffus) = self in
          (prefix=[] -> xiffus=[]) && view = prefix @ List.rev xiffus *)
\end{gospel}
It is worth mentioning that the first part of the invariant was already
expressed as an informal comment on the \textsf{OCamlGraph} sources. The
\texttt{view} field is used to represent the whole queue as a single list, with
elements in their right position. This is a ghost field since it has no
computational interest. We currently only support type invariants attached to
record definitions. However, after extraction, we recover the original
definition of type~\of{t} as an alias type.
In what follows, the specification of the operations on a queue is solely given
in terms of the \texttt{view} field. Let us begin with the simple operation of
creating a new, initially empty queue. Its implementation and specification are
the following:
\begin{gospel}
  let empty = { self = [], []; view = [] }
  (*@ t = empty
        ensures t.view = [] *)
\end{gospel}
When fed into \cameleer, a single VC is generated for the \texttt{empty}
function. This asserts that the freshly created queue respects the type
invariant (this is also the case for every operation returning a new
queue). This is easily discharged by any SMT solver.
The next function decides whether a given queue is empty. It is as simple as
follows:
\begin{gospel}
  let [@logic] is_empty (q: 'a t) = match q.self with
    | [], _ -> true
    | _ -> false
  (*@ b = is_empty q
        ensures b <-> q.view = [] *)
\end{gospel}
For this case, \why assumes the~\of{q} argument respects the queue invariant
(this is also the case for every operation taking a queue as argument). From a
computational point of view, a queue is empty if the prefix list is empty. From
a logical point of view, and according to the supplied type invariant, a queue
is empty if the \texttt{view} field is empty. The generated VC for the
\texttt{is\_empty} \post is easily discharged. We mark \texttt{is\_empty} with
the \texttt{logic} attribute in order to be able to use it within the
specification of subsequent operations. It is worth noting that
\texttt{is\_empty} is the only operation for which we (sightly) change the code,
compared to the one from \textsf{OCamlGraph}. The original implementation
amounts to the comparison \texttt{prefix = []}. The rationale to avoid such a
comparison is that \whyml does not support polymorphic comparison, other than
for integer values~\cite[Chap 12.2]{why3manual}. As future work, we plan on
extending \cameleer to support polymorphic equality for types for which we know
it is safe do so (\emph{e.g.}, non-cyclic lists).
The \texttt{add} function is the first one for which we must update the value of
the \texttt{view} model. This operation receives as arguments a \texttt{queue}
and a new element~\texttt{elt} to be inserted. The implementation is the
following:
\begin{gospel}
  let add queue elt = match queue.self with
    | [], [] ->
        { self = [elt], []; view = [elt] }
    | prefix, xiffus ->
        { self = prefix, elt :: xiffus; view = queue.view @ [elt] }
  (*@ r = add queue elt
        ensures r.view = queue.view @ [elt] *)
\end{gospel}
If the queue is empty, then~\texttt{elt} is pushed into the \texttt{prefix} list
and \texttt{view} is updated accordingly. Otherwise, the new element is added as
the new head of \texttt{xiffus} and as the last element of \texttt{view}. The
update on the \texttt{view} model should not jeopardize the constant-time
execution of \texttt{add}; this is a ghost expression, hence erased before
compilation time. \why generates VCs for type invariant preservation for the
newly created queues, as well as for the user-supplied \post. All of these are
automatically discharged in less than a second.
The last two operations we present are \texttt{head} and \texttt{tail}, used,
respectively, to retrieve the first element of the queue and to remove it. The
former is implemented and specified as follows:
\begin{gospel}
  let head queue = match queue.self with
    | [], _ -> raise Not_found
    | head :: _, _-> head
  (*@ x = head queue
        raises  Not_found -> is_empty queue
        ensures match queue.view with
                | [] -> false
                | y :: _ -> x = y *)
\end{gospel}
If the queue is empty we raise the \texttt{Not\_found} exception from the \ocaml
standard library. A \of{raises} clause is used to introduce what we call an
\emph{exceptional \post}. When the queue is not empty, we return the head
element of the \texttt{prefix} list. The given \post is valid since it stands
for the case when the function does not raise any exception. Hence, we have in
the proof context \texttt{not (is\_empty queue)} which allows one to prove the
\texttt{false} branch (in practice, this represents a case where there is a
logical contradiction on the proof context).
Finally, the \ocaml implementation and \GOSPEL specification of the
\texttt{tail} operation are as follows:
\begin{gospel}
  let [@logic] [@ghost] tail_list = function
      | [] -> assert false
      | _ :: l -> l
  (*@ r = tail_list param
        requires param <> []
        ensures  match param with
                 | [] -> false
                 | _ :: l -> r = l *)

  let tail q = match q.self with
    | [_], xiffus ->
        { self = (List.rev xiffus, []); view = tail_list q.view }
    | _ :: prefix, xiffus ->
        { self = (prefix, xiffus); view = tail_list q.view }
    | [], _ -> raise Not_found
  (*@ h = tail q
        raises  Not_found -> is_empty q
        ensures h.view = tail_list q.view *)
\end{gospel}
Similar to \texttt{head}, this function raises the \texttt{Not\_found} exception
whenever its argument is the empty queue. If~\texttt{q} is not empty, the above
implementation returns a new queue without the first element of~\of{q}. The
interesting case is when \texttt{prefix} is a singleton list. For such case, we
take the elements of \texttt{xiffus}, which are stored in reversed order, put
them back into the correct order and make this the new \texttt{prefix} list. We
use the \texttt{List.rev} from the \ocaml standard library, which is mapped by
\cameleer into the \texttt{reverse} function from the \why standard library. In
order to update the \texttt{view} field, we introduce the auxiliary
\texttt{tail\_list} that returns the tail of a non-empty list. Such a function
is used for proof and specification purposes only. All VCs generated for both
\texttt{tail\_list} and \texttt{tail} are automatically discharged in no time.
\myparagraph{Ephemeral implementation.} We present an alternative queue
implementation as an ephemeral data structure. Despite the underlying difference
of the two queue data type definitions, the \GOSPEL specification of the
ephemeral version operations is quite similar to the applicative ones. We
highlight here the most interesting differences between the implementations. In
particular we present an operation to concatenate two ephemeral queues, which is
not given in the \textsf{OCamlGraph} sources. The complete development of the
ephemeral queue can be found
online\footnote{\url{https://github.com/mariojppereira/cameleer/blob/master/examples/ephemeral_queue.ml}}.
}
Our first case study is the
implementation of a FIFO queue, implemented as an ephemeral data-structure. The
complete \ocaml development and \GOSPEL specification are presented at
\cameleer's GitHub repository
\footnote{\url{https://github.com/mariojppereira/cameleer/blob/master/examples/ephemeral_queue.ml}}.
We also publish online a simpler, purely applicative version of this data
structure
\footnote{\url{https://github.com/mariojppereira/cameleer/blob/master/examples/applicative_queue.ml}},
which we picked from the \textsf{OCamlGraph}
library\footnote{\url{https://github.com/backtracking/ocamlgraph}}.
This case study follows the standard approach of using a pair of lists to store
the elements of the queue, as follows:
\begin{gospel}
  type 'a t = { mutable front: 'a list; mutable rear : 'a list;
                mutable view : 'a list [@ghost]; }
 (*@ invariant (front = [] -> rear = []) && view = front ++ List.rev rear *)
\end{gospel}
In order to formally express the behavior of the queue data structure, we equip
type~\of{t} with a model field and a \GOSPEL invariant. The \texttt{view} field
is used to represent the whole queue as a single list. This is a ghost field
since it has no computational interest. Front elements are stored in the correct
order and rear elements are stored in \emph{reversed} order. Elements are pushed
into the head of the rear list and popped off the head of front. This data
structure also maintains the \emph{invariant} that if \texttt{front} is empty,
then so is \texttt{rear}.

In what follows, the specification of operations on a queue is solely given in
terms of the \texttt{view} field. The \texttt{push} operation is implemented and
specified as follows:
\begin{gospel}
  let push x q =
    if is_empty q then q.front <- [x] else q.rear <- x :: q.rear;
    q.view <- q.view @ [x]
  (*@ push x q
        ensures q.view = (old q.view) @ [x] *)
\end{gospel}
The \post asserts the updated \texttt{view} field of~\of{q} consists of the
value of \texttt{view} before the call (\texttt{old q.view}), extended with the
new element~\of{x}. If the queue is empty, then~\texttt{x} is pushed into the
\texttt{front} list; otherwise, the new element is added as the new head of
\texttt{rear}.

Next, we present the implementation of the \texttt{pop} operation. This is as
follows:
\begin{gospel}
  let pop q = match q.front with
    | [] -> raise Not_found
    | [x] ->
        q.front <- List.rev q.rear; q.rear <- []; q.view <- tail_list q.view;
        x
    | x :: f ->
        q.front <- f; q.view <- tail_list q.view;
        x
  (*@ x = pop q
        raises  Not_found -> is_empty (old q)
        ensures x :: q.view = (old q).view *)
\end{gospel}
If the queue is empty we raise the \texttt{Not\_found} exception from the \ocaml
standard library. A \of{raises} clause is used to introduce what we call an
\emph{exceptional \post}. We must be careful enough to specify that the
\texttt{is\_empty} property is verified by pre-state of~\of{q}. Otherwise, had
we used \of{is_empty q}, this would propagate into our proof context that the
queue is not empty \emph{after} the execution of \texttt{pop}. This would
prevent proving the safety of the next function.

The most interesting function in our development of ephemeral queues is the
concatenation of two such queues. The implementation is as follows:
\begin{gospel}
  let transfer (q1: 'a t) (q2: 'a t) : unit =
    while not (is_empty q1) do push (pop q1) q2 done
  (*@ transfer q1 q2
        raises   Not_found -> false
        ensures  q1.view = [] && q2.view = old q2.view @ old q1.view *)
\end{gospel}
The \texttt{transfer} operation takes queues~\of{q1} and~\of{q2} as arguments,
and migrates the elements of the former to the end of the latter. Moreover, it
clears the contents of its first argument. By design, \GOSPEL assumes~\of{q1}
and~\of{q2} are two separated queues, \emph{i.e.}, not aliased. The
\texttt{raises} clause states no \texttt{Not\_found} exception is raised during
execution of~\of{transfer}. In fact, since the \texttt{while} loop is guarded by
the \texttt{not (is\_empty q1)} property, we know it is safe to call \texttt{pop
  q1}. Feeding this program to \cameleer generates a total of 8 VCs, from which
we are only able to prove the exceptional \post and the first clause of the
regular \post condition. With no surprise, this comes from the fact that we are
missing a suitable loop invariant. In what follows, we refine the specification
to completely prove the \of{transfer} function.

In order to prove termination, we use the length of the \texttt{view} model
from~\of{q1} as a decreasing measure, as follows:
\begin{gospel}
    (*@ variant List.length q1.view *)
\end{gospel}
In order to prove the remaining VCs, it comes with no surprise that we need a
suitable loop invariant. First, we turn the type invariant property into a loop
invariant, as follows:
\begin{gospel}
    (*@ invariant (q1.front = [] -> q1.rear = []) &&
                  (q2.front = [] -> q2.rear = []) *)
    (*@ invariant q1.view = q1.front @ List.rev q1.rear &&
                  q2.view = q2.front @ List.rev q2.rear *)
\end{gospel}
Although this makes the loop invariant more cumbersome, it is actually almost a
mechanical process to include the type invariant in the loop invariant. With
such a specification, we are able to discharge every VC of \texttt{transfer},
except for the second \post.

To complete this proof, we need to be a little more creative. The idea is the
following: if we pick an arbitrary loop iteration, at that point of execution we
would have already transferred a prefix of the elements from~\of{q1} and would
have concatenated those into~\of{q2}. In order to represent such prefix
of~\of{q1}, we introduce an auxiliary ghost variable, as follows:
\begin{gospel}
  let transfer q1 q2 =
    let [@ghost] done_view = ref [] in ...
\end{gospel}
At each iteration, we must update the value of \texttt{done\_view} as follows:
\begin{gospel}
    while not (is_empty q1) do (*@ variant ... invariant ... *)
      done_view := !done_view @ [head_list q1.view];
      push (pop q1) q2
   done
\end{gospel}
Once again, the \texttt{done\_view} is a ghost variable, so any part of the
program that manipulates it should be erased from the code. Thus, the above
assignment should not incur a penalty on the execution time of
\texttt{transfer}. Finally, we complete the loop invariant as follows:
\begin{gospel}
      (*@ invariant old q1.view = !done_view @ q1.view *)
      (*@ invariant q2.view = old q2.view @ !done_view *)
\end{gospel}
The first condition maintains that \texttt{todo\_view} is indeed a prefix of
\of{q1.view}; the second condition states that the current state of~\of{q2}
consists of the elements from \of{todo_view} concatenated to the sequence of
original elements of~\of{q2}. With the updated invariant, we are finally able to
discharge every generated VC for \texttt{transfer}.

\ifthenelse{\boolean{longversion}}{
It is worth mentioning that reasoning about list segments is a common practice
when reasoning about ephemeral data structures. Such approach can easily become
complex and would require the use of heavier reasoning machinery (\emph{e.g.},
loop frame rule and magic wand from Separation Logic) which we currently do not
dispose from in \why or \GOSPEL~\cite{10.1145/3408998}. Nonetheless, \GOSPEL
allows us to write very concise and readable specification, to the level of what
an \ocaml programmer can understand. This makes proofs, such as the above
\texttt{transfer} function, still digestible.
}

\subsection{Leftist Heap}
\label{sec:leftist-heaps}

\paragraph{Functor definition.} The next case study is the implementation of a
\emph{heap} data structure. We adopt the \emph{leftist heap}
variant~\cite{10.5555/906397,knuth1998art}, which we picked from the
\textsf{OCaml-containers}
library\footnote{\url{https://github.com/c-cube/ocaml-containers}}. This is an
applicative implementation, following the approach by
C. Okasaki~\cite[Chap. 3.1]{okasaki1998}.

The fundamental interest of using heaps is to be able to quickly access the
minimum element of the collection. Thus, the elements of this data structure
must be equipped with a \texttt{total} preorder, which is crucial to guarantee
the correct behavior of the heap implementation. In \ocaml, such kind of
restrictions on types are naturally implemented using functors. For leftist
heaps, we begin by introducing the following \emph{module type} to represent a
total preorder:
\begin{gospel}
  module type TOTAL_PRE_ORD = sig
    type t
    (*@ function le : t -> t -> bool *)
    (*@ axiom reflexive : forall x. le x x *)
    (*@ axiom total     : forall x y. le x y \/ le y x *)
    (*@ axiom transitive: forall x y z. le x y -> le y z -> le x z *)

    val leq : t -> t -> bool
    (*@ b = leq x y
          ensures b <-> le x y *)
  end
\end{gospel}
Using \GOSPEL, we introduce a purely logical function~\of{le} defined using the
classic axioms of reflexivity, totality, and transitivity. Next, the
specification of the regular function \texttt{leq} should be read as ``this
function implements the logical function \texttt{le}''.
\ifthenelse{\boolean{longversion}}{ In other words, independently of its
  underlying implementation,~\texttt{leq} always returns the same result, for
  the same arguments, as function~\texttt{le}.  }
Using the above \texttt{TOTAL\_PRE\_ORD} type, the implementation of leftist
heaps is encapsulated in the following functor \texttt{Make}:
\begin{gospel}
  module Make(E : TOTAL_PRE_ORD) = struct type elt = E.t ... end
\end{gospel}
The type equation \texttt{elt = E.t} ensures that elements of type \texttt{elt}
(used to represent the elements of the heap) are inherit the total preorder
relation from module~\of{E}.

\myparagraph{Leftist property.} We use the following data type definition to
represent leftist heaps:
\begin{gospel}
    type t = E | N of int * elt * t * t
\end{gospel}
A leftist heap is represented as a binary tree where each node is attached a
\emph{rank} value. Generally speaking, the rank of a node is defined as the
length of the shortest path from that node to an empty node. In \GOSPEL, this is
as simple as follows:
\begin{gospel}
    (*@ function rank (h: t) : integer =
          match h with E -> 0 | N _ _ l r -> 1 + min (rank l) (rank r) *)
\end{gospel}
Using the general notion of rank, one can define the notion of \emph{leftist
  property}: the rank of any left child is always greater or equal to the rank
of the right sibling. This is captured by the following \GOSPEL definition:
\begin{gospel}
    (*@ predicate leftist (h: t) = match h with
          | E -> true
          | N n _ l r ->
              n = rank h && leftist l && leftist r && rank l >= rank r *)
\end{gospel}
This property also gives the value of the element storing the rank in the
structure. Other than the \texttt{leftist} property, leftist heaps should obey
the general laws of heaps: the element in each node is less or equal to the
elements at its children. The \GOSPEL definition is as follows:
\begin{gospel}
    (*@ predicate is_heap (h: t) = match h with
          | E -> true
          | N _ x l r -> le_root x l && is_heap l && le_root x r && is_heap r *)
\end{gospel}
where \texttt{le\_root} is a predicate that states a given element is less or
equal to the root of a heap:
\begin{gospel}
    (*@ predicate le_root (e: elt) (h: t) =
          match h with E -> true | N _ x _ _ -> E.le e x *)
\end{gospel}
Finally, we define what is a leftist heap:
\begin{gospel}
    (*@ predicate leftist_heap (h: t) = is_heap h && leftist h *)
\end{gospel}

\myparagraph{Logical definition of minimum element.} When describing the logical
behavior of a heap data structure, one must pay particular attention to the
definition of the minimum element.
\ifthenelse{\boolean{longversion}}{
Before proceeding into the
specification and verification of leftist heap functions, we introduce several,
important, auxiliary logical functions. First, we define the following
\texttt{size} function:
\begin{gospel}
    let [@logic] rec size = function
      | E -> 0
      | N (_,_ , l, r) -> 1 + size l + size r
    (*@ r = size param
          ensures 0 <= r
          ensures r = 0 <-> param = E *)
\end{gospel}
Here, we chose to introduce it as a \texttt{let [@logic]} definition in order to
be able to provide a function contract. Next, a \texttt{occ} function to count
the occurrences of a given element in the heap:
\begin{gospel}
    (*@ function occ (x: elt) (h: t) : integer = match h with
          | E -> 0
          | N _ e l r -> let occ_lr = occ x l + occ x r in
              if x = e then 1 + occ_lr else occ_lr *)

    (*@ predicate mem_heap (x: elt) (h: t) = 0 < occ x h *)
\end{gospel}
For such case, this is a \GOSPEL function since we are using polymorphic
equality, which is perfectly fine since this is a purely logical
definition. Last, but most important, we formally capture the notion of minimum
element of the heap.
}
\ifthenelse{\boolean{longversion}}{
First, we define the following relation:
\begin{gospel}
    (*@ predicate is_minimum (x: elt) (h: t) =
          occ x h > 0 && forall e. occ e h > 0 -> E.le x e *)
\end{gospel}
The logical function \of{occ} counts the number of occurrences of element~\of{x}
in a heap. It is straightforwardly defined as recursive traversal of
type~\of{t}. The element~\of{x} is a minimum element of~\of{h} if it belongs to
the tree and is less or equal to any other element of the heap. Nonetheless,
this definition is only an approximation: if the heap contains repeated
elements, then the minimum element must be any one that is the least according
to the \texttt{E.le} relation.}
We need to define what is the \emph{physical} minimum element of the heap, as
follows:
\begin{gospel}
    (*@ function minimum (h: t) : elt *)
    (*@ axiom minimum_def: forall l x r n. minimum (N n x l r) = x *)
\end{gospel}
The \texttt{minimum} function is only significantly defined for the case of
non-empty heaps\footnote{In \GOSPEL, logical functions are total. The value
  returned by \texttt{minimum E} exists, but one can prove no property about
  it.}.
\ifthenelse{\boolean{longversion}}
{
In order to state that the root of the heap is the minimum element, we
provide the following lemma:
\begin{gospel}
   let [@lemma] rec root_is_miminum = function
    | E -> assert false
    | N (_, _, l, r) ->
        begin match l with E -> () | _ -> root_is_miminum l end;
              match r with E -> () | _ -> root_is_miminum r
  (*@ root_is_minimum param
       requires is_heap param && param <> E
       ensures  is_minimum (minimum param) param
       variant  param *)
\end{gospel}
We define such a lemma as a \emph{lemma function}: an \ocaml ghost function,
whose contract is automatically turned into an axiom
afterwards~\cite{DBLP:conf/itp/Leino13}. We define such a lemma function as
recursive function, which allows us to encode a proof by induction of the
supplied \post.}
We have now the necessary building blocks to formally specify leftist heaps
operations.

\myparagraph{Heap operations.} We present here only two heap operations,
\of{_make_node} and \of{merge}. The complete \ocaml development can be found
online\footnote{\url{https://github.com/mariojppereira/cameleer/blob/master/examples/leftist_heap.ml}}. The
first function builds a new node, given an element~\of{x} and subtrees~\of{a}
and~\of{b}:
\begin{gospel}
  let _make_node x a b =
    if _rank a >= _rank b then N (_rank b + 1, x, a, b)
    else N (_rank a + 1, x, b, a)
  (*@ h = _make_node x a b
      requires leftist_heap a && leftist_heap b && le_root x a && le_root x b
      ensures  leftist_heap h && minimum h = x
      ensures  occ x h = 1 + occ x a + occ x b
      ensures  forall y. x <> y -> occ y h = occ y a + occ y b *)
\end{gospel}
The leftist property is ensured by the \texttt{if..then..else} expression, where
\texttt{\_rank} is a function that simply retrieves the rank of a node,
\emph{i.e.}, the first argument of the~\texttt{N} constructor or~$0$ in case the
heap is empty. Both~\of{a} and~\of{b} must be \texttt{leftist\_heap}s and~\of{x}
must no greater than the roots of the two arguments. We give a specification of
the heap in terms of the multiset of its elements. The resulting heap~\of{h} is
a \texttt{leftist\_heap}, with minimum element~\of{x}. The number of occurrences
of~\of{x} increases by one, whereas the occurrences of any element different
from~\of{x} remain the same. Finally, the nuclear operation \texttt{merge} is
defined as follows:
\begin{gospel}
    let rec merge t1 t2 = match t1, t2 with
      | t, E | E, t -> t
      | N (_, x, a1, b1), N (_, y, a2, b2) ->
          if E.leq x y then _make_node x a1 (merge b1 t2)
          else _make_node y a2 (merge t1 b2)
    (*@ h = merge t1 t2
          requires leftist_heap t1 && leftist_heap t2
          variant  size t1 + size t2
          ensures  leftist_heap t && forall x. occ x h = occ x t1 + occ x t2 *)
\end{gospel}
If the root~\of{x} of heap~\of{t1} is no greater than the root of~\of{t1}, we
build a new heap with root~\of{x}; otherwise, we the new root is the root
of~\of{t2}. The call to function \texttt{\_make\_node} ensures the
\texttt{leftist\_heap} property for the returned heap. We use the logical
function \texttt{size} in the variant of \texttt{merge}, which is the
straightforward definition of the number of nodes in a heap.

Addition and removal of the minimal element are straightforwardly defined. Due
to space constraints, we do not present those here and refer the reader to the
online repository for the complete \ocaml development. This also includes the
\of{filter} and \of{delete_all} higher-order functions.
\ifthenelse{\boolean{longversion}}
{We add new elements to the heap using the \texttt{insert} function and remove
the minimum element using \of{find_min_exn}.
\begin{gospel}
    let insert x h = merge (N (1, x, E, E)) h
    (*@ new_h = insert x h
          requires leftist_heap h
          ensures  leftist_heap new_h && occ x new_h = 1 + occ x h
          ensures  forall y. x <> y -> occ y new_h = occ y h *)
\end{gospel}
and remove then via the following \texttt{find\_min\_exn} operation:
\begin{gospel}
    let find_min_exn = function
      | E -> raise Empty
      | N (_, x, _, _) -> x
    (*@ r = find_min_exn param
          requires leftist_heap param
          raises   Empty -> param = E
          ensures  r = minimum param *)
\end{gospel}
This function returns the minimum element of the heap, without altering the data
structure; on the other hand, if the heap is empty, it raises the \texttt{Empty}
exception.}
\ifthenelse{\boolean{longversion}}{Lastly, we present the \texttt{filter} function, used to keep in the collection
only the elements that satisfy a given predicate. Its \ocaml implementation and
\GOSPEL specification are as follows:
\begin{gospel}
    let rec filter p = function
      | E -> E
      | N (_, x, l, r) ->
          if p x then _make_node x (filter p l) (filter p r)
          else merge (filter p l) (filter p r)
    (*@ h = filter p param
          requires leftist_heap param
          variant  param
          ensures  leftist_heap h
          ensures  forall x. not (p x) -> occ x h = 0
          ensures  forall x. p x -> occ x h = occ x param *)
\end{gospel}
This is our first encounter with a higher-order definition. In \GOSPEL, we
assume every first-class function to be
effect-free~\cite[Sec. 2.3]{DBLP:conf/fm/ChargueraudFLP19}. For the above
example, this allows us to treat~\of{p} directly in the specification as logical
function. Other than the \texttt{leftist\_heap} property, the \post of
\texttt{filter} ensures that every element different from~\of{x} is removed from
the final heap and that no new elements are added to the collection.}
All the generated VCs for leftist heap operations were automatically discharged.

\ifthenelse{\boolean{longversion}} { No manual assistance was needed in order to
  complete this proof, other than very simple calls to the \why
  \texttt{split\_vc} transformation.  }

\vspace{-3pt}
\subsection{Other case studies}
\label{sec:other-case-studies}


Table~\ref{tab:case-studies} in Appendix~\ref{sec:summ-camel-case} summarizes
the case studies performed with \cameleer. The second column features the number
of non-blank lines of \ocaml code, while the third one stands for the number of
non-blank lines of \GOSPEL specification. We have put an effort to use \cameleer
to verify programs of different natures. These include numerical programs
(binary multiplication, different factorial implementations, fast
exponentiation, and integer square root), sorting and searching algorithms, data
structures implemented as functors, historical algorithms (checking a large
routine by Turing, Boyer-Moore's majority algorithm, and binary tree same
fringe), and logical algorithms (conversion of a propositional formula into
conjunctive normal form).

\mysection{Related Work}
\label{sec:related-work}

\ifthenelse{\boolean{longversion}}{
\myparagraph{Deductive program verification.} A growing interest from
industry, as well as the results of an active research community, have made it
possible to apply deductive tools and principles in the verification of large,
realistic software projects. The Sel4 project~\cite{sel4} presents the formal
verification of a complete, general-purpose operating system kernel. The proof
is conducted using the Isabelle/HOL theorem
prover~\cite{nipkow2002isabelle}. CompCert~\cite{leroy2009formal} and
CakeML~\cite{DBLP:journals/jfp/TanMKFON19} are realistic, optimizing, efficient
compilers for the C language and an ML-like dialect (respectively), whose
functional correctness was verified using proof assistants. On the other end of
the spectrum, a remarkable application of automated verification tools, the
Verisoft XT project~\cite{BeckertMoskal2010}, introduced a verified version of
Microsoft's Hypervisor and its embedded operating system PikeOS. The proof is
done using VCC~\cite{vcc} and the Z3~\cite{z3} SMT solver. More recently,
Clochard \textit{et al.}~\cite{10.1145/3371070} presented a new approach for the
automated verification of programs based on the notion of%
\emph{ghost monitors}. Such a technique has been successfully applied in the
verification of a symbolic interpreter
engine~\cite{DBLP:conf/vstte/BeckerM19}. Our aim is to develop and strengthen
\cameleer up to a mature tool that can be used in the verification of
industrial-size software, following the cited examples.
}

\paragraph{Automated deductive verification tools.}
\ifthenelse{\boolean{longversion}}{ Over the last decades, the development of
  automated verification tools has made significant progress. The so-called
  \emph{SMT revolution} and the development of reusable intermediate
  verification infrastructures contributed decisively to the development of
  practical automated deductive verifiers.}

One can cite \why, \textsf{F*}~\cite{dm4free},
\dafny~\cite{10.1007/978-3-642-17511-4-20}, and
\viper~\cite{DBLP:conf/vmcai/0001SS16} as well-succeed automated deductive
verification tools. Formal proofs are conducted in the proof-aware language of
these frameworks, and then executable reliable code can be automatically
extracted. In the \cameleer project, we chose to develop a verification tool
that accepts as input a program written directly in \ocaml, instead of a
dedicated proof language. Our specification language, \GOSPEL, is very close to
the \ocaml language itself, hence we believe this does not impose a big burden
for the regular \ocaml practitioner.

Regarding verification tools that tackle the verification of programs written in
mainstream languages, we can cite
Frama-C~\cite{DBLP:journals/fac/KirchnerKPSY15} and
VeriFast~\cite{DBLP:conf/nfm/JacobsSPVPP11}. The former is a framework for the
static analysis of C code; the latter can be used to verify functional
correctness of C and Java programs. Despite the remarkable case studies verified
with these tools, C and Java code can quickly degenerate into a nightmare of
pointer manipulation and tricky semantic issues. We argue \ocaml is a language
better suited for formal verification.

\myparagraph{Deductive verification of \ocaml programs.} To the best of our
knowledge, \cfml~\cite{chargueraud2011characteristic} was the only tool
available for the deductive verification of code directly written in \ocaml. It
takes as input an \ocaml program and translates it into \coq, together with its
\emph{characteristic formulae}. A characteristic formulae is a higher-order
statement that captures the semantics of the original program. Proofs are
conducted using an embedding of Separation Logic inside \coq. The \cfml tool has
already been used to verify several interesting \ocaml modules, including
ephemeral data structures and higher-order functions. Recently, \cfml was
extended with support for \emph{time credits} and it was successfully applied in
the verification of functional correctness and time complexity claims of
non-trivial data structures and
algorithms~\cite{chargueraud2019verifying,DBLP:conf/itp/GueneauJCP19}.

The \vocal project aims at developing a mechanically verified \ocaml
library~\cite{chargueraud:hal-01561094}. One of the main novelties of this
project is the combined use of three different verification tools: \why, \cfml,
and \coq. The \GOSPEL specification language was developed in the scope of this
project, as a tool-agnostic language that could be manipulated by any of the
three mentioned frameworks.
\ifthenelse{\boolean{longversion}}{
For instance, if the code is purely functional, it
can be verified directly in \coq; if it features higher-order effectful
computations, then \cfml is the tool of choice; finally, if mutability is
limited, \why can be used to achieve greater degree of proof automation.
}
It was our participation on the \vocal project that inspired us, in the first
place, to develop \cameleer. We believe \cameleer can be readily included in the
\vocal ecosystem, complementing the toolchains~\cite{filliatre:hal-01783851}
that are currently used to build the verified library.
\ifthenelse{\boolean{longversion}}{
We believe that \cameleer can not only be a fine addition to the \vocal
ecosystem, but it can also improve the existing
toolchains~\cite{filliatre:hal-01783851}. With \why, one builds verified
implementations in \whyml and extract executable \ocaml code afterwards; with
\cameleer, we directly verify existing \ocaml code, while using \why on the
background for proof automation.
}

\mysection{Conclusions and Future Work}
\label{sec:conclusion}

In this paper we presented \cameleer, a tool for deductive verification of
\ocaml-written code.
\ifthenelse{\boolean{longversion}}{
We believe \ocaml is an excellent target for formal software verification, and
it is actually surprising that little attention (with some notable exceptions)
has been given to \ocaml by the formal methods community. With the goal in mind
of bringing more \ocaml programmers to adopt a verification tools in their
development routines, we have built \cameleer with a clear focus towards
automated proof. Proof assistant tools, which demand a lot of human effort and
high degree of expertise, are likely to become a \textit{de facto} choice by the
lambda-\ocaml programmer who is not a proof expert. Moreover, we use the \GOSPEL
language to provide rigorous, yet readable, behavioral specification to \ocaml
code.
}
The core of \cameleer is a translation from \ocaml annotated code into \whyml,
the programming and specification language of the \why verification
framework. \ocaml and \whyml have many common traits (both in their syntax and
semantics), which provides us with good guarantees about soundness of \cameleer
translation. We have already applied \cameleer to successfully verify functional
correctness and safety of 14 realistic \ocaml modules. These include
implementations issued from existing libraries, and scale up to data structures
implemented as functors and tricky effectful computations.
\ifthenelse{\boolean{longversion}}
{
Even if the proof of
each module is ultimately achieved via a combination of SMT solvers, we still
need some iteration human interaction, namely writing the \GOSPEL specification
and manipulating ghost models. Nonetheless, we are convinced such effort is less
scary for non-experts than the use of a interactive proof assistant, such as
\coq.
}
These results encourage us to continue developing \cameleer and apply it to the
verification of larger case studies.

\myparagraph{What we do not support.} Currently, we target a subset of the
\ocaml language which roughly corresponds to \texttt{caml-light}, with basic
support for the module language (including functors). Also, \whyml limits
effectful computations to the cases where alias is statically known, which
limits our support for higher-order functions and mutable recursive data
structures. Adding support for the objective layer of the \ocaml language would
require a major extension to the \GOSPEL language and a redesign of our
translation into \whyml. Nonetheless, \why has been used in the past to verify
Java-written programs~\cite{DBLP:conf/cav/FilliatreM07}, so in principle an
encoding of \ocaml objects in \whyml is possible.

GADTs extend usual algebraic data types with a lightweight form of dependent
typing in \ocaml. The use of GADTs allows one to logically constraint the values
that can inhabit a type in a certain point of the program. The \whyml type
system does not, currently, include a notion close to GADTs. However, since this
a proof-aware language an interesting route for future work would be to explore
if it is possible to encode in \whyml the reasoning of GADTs without extending
its type system. In particular, if it would be possible to leverage on the
notion of type invariant to achieve similar results as those statically provided
by GADTs.

Another interesting feature of \ocaml we currently do not support are
polymorphic variants. Polymorphic variants are more flexible than ordinary
variants, as they are not tied to a particular type declaration and can be
easily extended according to use scenarios. This flexibility, however, leads to
a more complicated typing process for polymorphic variants, when compared to
regular ones. Once again, extending \cameleer to deal with polymorphic variants
requires extending \why itself. This would likely mean a considerable redesign
of its type system.

Finally, \why constrains the use of higher-order functions to pure computations
and this is also the case with \cameleer. A possible solution for such
limitation would be to use defunctionalization to convert an higher-order
program into an equivalent first-order one. In previous work, we have already
explored the use of defunctionalization for verification of stateful
higher-order programs in \why~\cite{pereira17jfla}. However, defunctionalization
is a whole-program transformation which severally constrains its
applicability. Next, we describe a more robust solution, which amounts at using
a richer verification framework to reason about higher-order with effects.

\ifthenelse{\boolean{longversion}}{
\myparagraph{Who verifies the verifiers.} One classical question posed to the
formal methods community is how reliable are the verification tools we
use. Could it be the case that a bug on the implementation of such tools would
allow one to prove an inconsistent result? In other words, are verification
tools implemented on top of verified software? With some remarkable
exceptions~\cite{Jacobs_2015,DBLP:conf/popl/JourdanLBLP15}, the answer is most
of the times no. As a much long-term project, we would like to evolve \cameleer
to the point where it would be able to cope with the proof of parts of
verification tools, \emph{e.g.}, from the \why core and the \cameleer tool
itself. In the case of \cameleer, an interesting task would be to prove that the
\ocaml to \whyml preserves the semantics and typedness of the original program.

\myparagraph{A more robust toolchain.} We currently rely on a patched version of
the \ocaml compiler libraries, provided by the \GOSPEL toolset, to build our
\ocaml to \whyml translation. Relying on the \texttt{compiler-libs} API is a
common problem within the \ocaml community, mainly for compatibly issues when a
new version of the compiler is released. Our short-term plan is to adapt our
implementation to use more robust solutions, such as the \texttt{ppxlib}
library\footnote{\url{https://github.com/ocaml-ppx/ppxlib}}. In fact, this is
already work in progress, by other members of the \vocal project, to remove the
dependency of \GOSPEL on the compiler
code\footnote{\url{https://github.com/vocal-project/vocal/pull/23}}.}

\myparagraph{Interface with \viper and \cfml.} We want to keep improving
\cameleer with the ability to verify a growing class of \ocaml
implementations. This includes pointer-based data structures and effectful
higher-order computations. Given the limitations of \why to deal with such class
of programs, we believe the solution is to extend \cameleer to include
translation into different intermediate verification languages. We are
considering targeting the \viper infrastructure and the \cfml tool. On one hand,
\viper is an intermediate verification language based on Separation Logic but
oriented towards SMT-based software verification, allowing one to automatically
verify heap-dependent programs. On the other hand, the \cfml tool already
provides a translation from \ocaml into \coq, allowing one to verify effectful
higher-order functions. Even if it relies on an interactive proof assistant,
\cfml provides a comprehensive library of tactics that ease the proof
effort. Our ultimate goal is to turn \cameleer into a verification tool that can
simultaneously benefit from the best features of different verification
frameworks. Our motto: we want \cameleer to be able to verify parts of an \ocaml
module using \why, others with \viper, and finally some specific functions with
\cfml.

\paragraph{Acknowledgements.} We thank Jean-Christophe Filliâtre and the
anonymous reviewers for feedback on previous versions of this paper. We also
thank Simon Cruanes for encouraging us to prove the Leftist Heaps implementation
from the \texttt{ocaml-containers} library.

\bibliographystyle{plain}

\newpage
\appendix

\section{Translation of expressions}
\label{sec:transl-expr}

\begin{figure}[h]
  \centering
\begin{judge}[b]{\textwidth}\footnotesize
  \[
    \begin{array}{c}
      \inferrule*[Left=(EAbsurd)]
      { }
      {\translate{\mathtt{assert\;false}}{expression}{\mathtt{absurd}}} \\[1em]

      \inferrule*[Left=(EFun)]
      {\mathtt{ghost}(\atts) = \beta \\
        \translate{\atts}{\textit{function spec}}{\mathcal{S}} \\
        \translate{e}{expression}{e'}}
      {\translate{\mathtt{fun}\;\atts\;x\rightarrow e}{expression}
      {\mathtt{fun}\;(\beta x)\;\mathcal{S} = e'}}\qquad

      \inferrule*[Right=(EApp)]
      {\translate{\overline{e}}{expression}{\overline{e'}}}
      {\translate{f\,\expbar}{expression}{f\,\overline{e'}}}\\[1em]

      \inferrule*[Lab=(ERec)]
      {
       \neg\textit{is\_ghost}(\atts_0) \\
       \mathit{kind}(\atts_0) = \kind \\
       \translate{\atts_1}{\textit{function spec}}{\spec_1} \\
       \translate{e_0}{function}{\overline{x\beta},e_0'} \\\\
       \translate{\overline{e_1}}{function}{\overline{\beta y},\overline{e_1'}} \\
       \translate{\overline{\atts_2}}{\textit{function spec}}{\overline{\spec_2}} \\
       \translate{e_2}{expression}{e_2'}}
      {\translate{\mathtt{let} \: \atts_0 \; \mathtt{rec} \: f_0 = e_0 \;
       \atts_1 \; \overline{\mathtt{and}\;f_1 = e_1 \; \atts_2} \;
       \mathtt{in} \; e_2}{expression}
       {\mathtt{rec}\;\kind\;f(\overline{\beta x})\;\spec_1 = e_0' \;
        \overline{\mathtt{with}\;\kind\;f_1 (\overline{\beta y}) = e_1' \;
        \spec_2}\;\mathtt{in} \; e_2'}}\\[1em]

      \inferrule*[Lab=(ERecGhost)]
      {\textit{is\_ghost}(\atts_0) \\
       \mathit{kind}(\atts_0) = \kind \\
       \translate{\atts_1}{\textit{function spec}}{\spec_1} \\
       \translate{e_0}{function}{\overline{x\beta},e_0'} \\\\
       \translate{\overline{e_1}}{function}{\overline{\beta y},\overline{e_1'}} \\
       \translate{\overline{\atts_2}}{\textit{function spec}}{\overline{\spec_2}} \\
       \translate{e_2}{expression}{e_2'}}
      {\translate{\mathtt{let} \: \atts_0 \; \mathtt{rec} \: f_0 = e_0 \;
       \atts_1 \; \overline{\mathtt{and}\;f_1 = e_1 \; \atts_2} \;
       \mathtt{in} \; e_2}{expression}
       {\mathtt{rec}\;\kind\;f(\overline{\beta x})\;\spec_1 = \texttt{ghost}\:e_0'
      \;\overline{\mathtt{with}\;\kind\;f_1 (\overline{\beta y}) = e_1' \;
        \spec_2}\;\mathtt{in} \; e_2'}}\\[1em]


      \inferrule*[Left=(ELet)]
      {\neg\textit{is\_ghost}(\atts) \\ \mathtt{kind}(\atts) = \kind \\
      \neg\textit{is\_functional}(e_0) \\\\
      \translate{e_0}{expression}{e_0'} \\ \translate{e_1}{expression}{e_1'}}
      {\translate{\mathtt{let}\:\atts\:x = e_0\;\atts'\;\mathtt{in}\;e_1}
      {expression}{\mathtt{let}\: \kind \: x = e_0'\;\mathtt{in} \;e_1'}}\\[1em]

      \inferrule*[Left=(ELetGhost)]
      { \textit{is\_ghost}(\atts) \\ \mathtt{kind}(\atts) = \kind \\
      \neg\textit{is\_functional}(e_0) \\\\
      \translate{e_0}{expression}{e_0'} \\ \translate{e_1}{expression}{e_1'}}
      {\translate{\mathtt{let}\:\atts\:x = e_0\;\atts'\;\mathtt{in}\;e_1}
      {expression}{\mathtt{let}\: \kind \: x = \mathtt{ghost}\:e_0'\;
      \mathtt{in} \;e_1'}} \\[1em]

      \inferrule*[Left=(ELetGhostFun)]
      { \textit{is\_ghost}(\atts) \\ \textit{kind}(\atts) = \kind \\
      \textit{is\_functional}(e_0) \\\\
      \translate{\atts'}{\textit{function spec}}{\spec} \\
      \translate{e_0}{function}{\betaxbar, e_0'} \\
      \translate{e_1}{expression}{e_1'}}
      {\translate{\mathtt{let}\:\atts\:f = e_0\;\atts'\;\mathtt{in}\;e_1}
      {expression}{\mathtt{let}\;\kind\;f =
      \mathtt{fun}(\betaxbar)\:\spec\rightarrow
      \mathtt{ghost}\:e_0'\;\mathtt{in} \;e_1'}} \\[1em]

      \inferrule*[Left=(ELetFun)]
      { \neg\textit{is\_ghost}(\atts) \\ \textit{kind}(\atts) = \kind \\
      \textit{is\_functional}(e_0) \\\\
      \translate{\atts'}{\textit{function spec}}{\spec} \\
      \translate{e_0}{function}{\overline{x\beta}, e_0'} \\
      \translate{e_1}{expression}{e_1'}}
      {\translate{\mathtt{let}\:\atts\:f = e_0\;\atts'\;\mathtt{in}\;e_1}
      {expression}{\mathtt{let}\;\kind\;f =
      \mathtt{fun}(\overline{\beta x})\:\spec\rightarrow e_0'\;\mathtt{in} \;e_1'}
      }\\[1em]

      \inferrule*[Left=(EWhile)]
      {\translate{\atts}{\textit{loop annotation}}{\loops} \\
      \translate{e_0}{expression}{e_0'} \\ \translate{e_1}{expression}{e_1'}}
      {\translate{\mathtt{while} \: e_0 \: \mathtt{do} \: \atts \: e_1 \:
      \mathtt{done}}{expression}
      {\mathtt{while} \: e_0' \: \mathtt{do} \: \loops \: e_1' \:
      \mathtt{done}}} \\[1em]

      \inferrule*[Left=(EIf)]
      {\translate{e_0}{expression}{e_0'} \\ \translate{e_1}{expression}{e_1'} \\
      \translate{e_2}{expression}{e_2'}}
      {\translate{\mathtt{if}\: e_0\: \mathtt{then}\: e_1\: \mathtt{else}\: e_2}
      {expression}
      {\mathtt{if}\: e_0'\: \mathtt{then}\: e_1'\: \mathtt{else}\: e_2'}}
      \\[1em]

      \inferrule*[Left=(EMatch)]
      {\translate{e_0}{expression}{e_0'} \\
      \translate{\overline{e_1}}{expression}{\overline{e_1'}}}
      {\translate{\mathtt{match}\; e_0\;
      \mathtt{with}\;\overline{\talloblong\,p\Rightarrow e_1}}
      {expression}
      {\mathtt{match}\; e_0'\;
      \mathtt{with}\;\overline{\talloblong\,p\Rightarrow e_1'}\,\texttt{end}}}

      \\[1em]

      \inferrule*[Left=(ERaise)]
      {\translate{\expbar}{expression}{\expbar'}}
      {\translate{\mathtt{raise} \; E\expbar}{expression}{\mathtt{raise} \;
        E\expbar'}}\qquad

      \inferrule*[Right=(ETry)]
      {\translate{e}{expression}{e'} \\ \translate{\exbar}{expression}{\exbar'}}
      {\translate{\mathtt{try} \; e \; \mathtt{with} \; \exbar}{expression}
      {\mathtt{try} \; e' \; \mathtt{with} \; \exbar'}\texttt{end}}
    \end{array}
  \]
\end{judge}
  \caption{Translation of \ocaml expressions into \whyml.}
  \label{fig:expressions}
\end{figure}

\newpage

\section{Translation of top-level declarations and type definitions}
\label{sec:transl-top-level}

\begin{figure}[h!]
  \centering
  \begin{judge}[b]{\textwidth}\footnotesize
    \[
      \begin{array}{c}
        \inferrule*[Left=(DModule)]
        {\translate{m}{module}{\overline{d}}}
        {\translate{\mathtt{module}\:\mathcal{M} = m}{declaration}
        {\mathtt{scope}\:\mathcal{M}\;\overline{d}\;\mathtt{end}}} \\[1em]

        \inferrule*[Left=(DLetGhost)]
        {\mathit{is\_ghost}(\atts) \\ \textit{kind}(\atts') = \mathcal{K} \\
        \textit{is\_functional}(e)\\\\
        \translate{\atts}{\textit{function spec}}{\mathcal{S}} \\
        \translate{e}{function}{\betaxbar, e'}}
        {\translate{\mathtt{let}\: \atts \; f = e \: \atts'}{declaration}
        {\mathtt{let}\:f\:\mathcal{K} =
        \mathtt{fun}(\betaxbar)\:\mathcal{S}\rightarrow
        \mathtt{ghost}\;e'}}\\[1em]

        \inferrule*[Left=(DLet)]
        {\neg\mathit{is\_ghost}(\atts) \\ \textit{kind}(\atts') = \mathcal{K} \\
        \textit{is\_functional}(e)\\\\
        \translate{\atts}{\textit{function spec}}{\mathcal{S}} \\
        \translate{e}{function}{\betaxbar, e'}
        }
        {\translate{\mathtt{let}\: \atts \; f = e \: \atts'}{declaration}
        {\mathtt{let}\:f\:\mathcal{K} =
        \mathtt{fun}(\betaxbar)\:\mathcal{S}\rightarrow\;e'}} \\[1em]

      \inferrule*[Lab=(DRec)]
      {
       \neg\textit{is\_ghost}(\atts_0) \\
       \mathit{kind}(\atts_0) = \kind \\
       \translate{\atts_1}{\textit{function spec}}{\spec_1} \\
       \translate{e_0}{function}{\overline{x\beta},e_0'} \\\\
       \translate{\overline{e_1}}{function}{\overline{y:\pi},\overline{e_1'}} \\
       \translate{\overline{\atts_2}}{\textit{function spec}}{\overline{\spec_2}}}
      {\translate{\mathtt{let} \: \atts_0 \; \mathtt{rec} \: f_0 = e_0 \;
       \atts_1 \; \overline{\mathtt{and}\;f_1 = e_1 \; \atts_2}}{declaration}
       {\mathtt{rec}\;\kind\;f(\overline{\beta x})\;\spec_1 = e_0' \;
        \overline{\mathtt{with}\;\kind\;f_1 (\overline{y:\pi}) = e_1' \;
        \spec_2}}}\\[1em]

      \inferrule*[Lab=(DRecGhost)]
      {\textit{is\_ghost}(\atts_0) \\
       \mathit{kind}(\atts_0) = \kind \\
       \translate{\atts_1}{\textit{function spec}}{\spec_1} \\
       \translate{e_0}{function}{\overline{x\beta},e_0'} \\\\
       \translate{\overline{e_1}}{function}{\overline{y:\pi},\overline{e_1'}} \\
       \translate{\overline{\atts_2}}{\textit{function spec}}{\overline{\spec_2}}}
      {\translate{\mathtt{let} \: \atts_0 \; \mathtt{rec} \: f_0 = e_0 \;
       \atts_1 \; \overline{\mathtt{and}\;f_1 = e_1 \; \atts_2}}{declaration}
       {\mathtt{rec}\;\kind\;f(\overline{\beta x})\;\spec_1 = \texttt{ghost}\:e_0'
      \;\overline{\mathtt{with}\;\kind\;f_1 (\overline{y:\pi}) = e_1' \;
        \spec_2}}}\\[1em]



        \inferrule*[Left=(DType)]
        {\translate{td_0}{\textit{type definition}}{td_0'} \\
         \translate{\overline{td_1}}
         {\textit{type definition}}{\overline{td_1'}}}
        {\translate{\texttt{type} \; td_0  \; \overline{\texttt{and}\; td_1}}
         {declaration}
         {\texttt{type}\: td_0' \; \overline{\texttt{with}\:td_1'}}}\\[1em]

        \inferrule*[Left=(DExn)]
        { }
        {\translate{\texttt{exception}\: E : \pibar}{declaration}{
        \texttt{exception}\: E : \pibar}}
      \end{array}
    \]
  \end{judge}
  \caption{Translation of \ocaml top-level declarations into \whyml.}
  \label{fig:top-level}
\end{figure}

\begin{figure}
  \centering
  \begin{judge}[b]{\textwidth}\small
    \[
      \begin{array}{c}
        \inferrule*[Left=(TDAbstract)]
        { }
        {\translate{\alphabar\:T}{\textit{type definition}}{T\alphabar}}\qquad~

        \inferrule*[Right=(TDAlias)]
        { }
        {\translate{\alphabar\:T = \tau}{\textit{type definition}}
         {T\alphabar = \tau}}\\[1em]

        \inferrule*[Left=(TDRecord)]
        {\translate{\atts}{\textit{type invariant}}{\bar{t}}}
        {\translate{\alphabar\:T = \{\: \overline{f : \pi}\:\}\:\atts}
        {declaration}
        {T\alphabar = \{\:\overline{f:\pi}\:\}\:\mathtt{invariant}\:\bar{t}}}
        \\[1em]

        \inferrule*[Left=(TDVariant)]
        { }
        {\translate{\alphabar\:T =
         \overline{\talloblong\,C\,\texttt{of}\,\overline{\tau}}}
         {\textit{type definition}}
         {T\alphabar = \overline{\talloblong\,C\,\overline{\tau}}}} \\[1em]
      \end{array}
    \]
  \end{judge}
  \caption{Translation of \ocaml type definitions into \whyml.}
  \label{fig:type-def}
\end{figure}

\newpage

\section{Translation of module expressions and module types}
\label{sec:transl-module-expr}

\begin{figure}[h!]
  \centering
  \begin{judge}[b]{\textwidth}\small
  \[
    \begin{array}{c}
      \inferrule*[Left=(MStruct)]
      {\translate{\overline{d}}{declaration}{\overline{d'}}}
      {\translate{\mathtt{struct}\:\overline{d}\:\mathtt{end}}{\textit{module}}
      {\overline{d'}}} \\[1em]

      \inferrule*[Left=(MFunctor)]
      {\translate{mt}{\textit{module type}}{\overline{d}} \\
      \translate{m}{module}{\overline{d'}}}
      {\translate{\mathtt{functor}(\mathcal{X}: mt)\rightarrow m}
      {\mathit{module}}
      {\mathtt{scope}\:\mathcal{X}\;\overline{d}\;\mathtt{end}\;\overline{d'}}}
      \\[1em]

      \inferrule*[Left=(MTSig)]
      {\translate{\overline{s}}{signature}{\overline{d}}}
      {\translate{\mathtt{sig}\:s\:\mathtt{end}}{\textit{module type}}
      {\overline{d}}}
    \end{array}
  \]
  \end{judge}
  \caption{Translation of \ocaml module expressions and module types into
    \whyml.}
  \label{fig:mod}
\end{figure}

\newpage

\section{Translation of signatures}
\label{sec:transl-sign}

\begin{figure}[h!]
  \centering
  \begin{judge}[b]{\textwidth}\small
  \[
    \begin{array}{c}
      \inferrule*[Left=(SVal)]
      {\neg\textit{is\_ghost}(\atts) \\ \textit{kind}(\atts) = \kind \\\\
      \translate{\atts'}{\textit{function spec}}{\spec} \\
      \translate{\pi,\atts'}{\textit{function args}}{\overline{x:\pi'},\pi_{res}}}
      {\translate{\mathtt{val}\: \atts \: f : \pi \: \atts'}{signature}
      {\mathtt{val}\:\kind\: f (\overline{x:\pi})\:\spec : \pi_{res}}}\\[1em]

      \inferrule*[Left=(SValGhost)]
      {\neg\textit{is\_ghost}(\atts) \\ \textit{kind}(\atts) = \kind \\\\
      \translate{\atts'}{\textit{function spec}}{\spec} \\
      \translate{\pi,\atts'}{\textit{function args}}{\overline{x:\pi'},\pi_{res}}}
      {\translate{\mathtt{val}\: \atts \: f : \pi \: \atts'}{signature}
      {\mathtt{val}\:\kind\:\ghost f (\overline{x:\pi})\:\spec : \pi_{res}}}\\[1em]

      \inferrule*[Left=(SType)]
      {\translate{td_0}{\textit{type definition}}{td_0'} \\
      \translate{\overline{td_1}}
      {\textit{type definition}}{\overline{td_1'}}}
      {\translate{\texttt{type} \; td_0  \; \overline{\texttt{and}\; td_1}}
      {signature}
      {\texttt{type}\: td_0' \; \overline{\texttt{with}\:td_1'}}}
    \end{array}
  \]
  \end{judge}
  \caption{Translation of \ocaml signatures into \whyml.}
  \label{fig:sigs}
\end{figure}

\newpage

\section{Summary of \cameleer case studies}
\label{sec:summ-camel-case}

\begin{table}[h!]
  \centering
  \begin{tabular}{lrr}
    Case Study               & Lines of Code & Lines of Specification\\
    \hline\hline
    Applicative Queue        & 25  & 21 \\
    Binary Multiplication    & 10  & 6  \\
    Binary Search            & 15  & 10 \\
    Checking a Large Routine & 25  & 15 \\
    CNF Convertion           & 113 & 47 \\
    Ephemeral Queue          & 39  & 32 \\
    Factorial                & 10  & 13 \\
    Fast Exponentiation      & 19  & 47 \\
    Fibonacci                & 46  & 58 \\
    Insertion Sort           & 13  & 80 \\
    Integer Square Root      & 8   & 14 \\
    Leftist Heap             & 68  & 108 \\
    Mjrty                    & 45  & 32 \\
    Same Fringe              & 23  & 16 \\\hline
    total                    & 459 & 499
  \end{tabular}
  \caption{Case studies verified with the \cameleer tool.}
  \label{tab:case-studies}
\end{table}

\end{document}